\documentclass[sn-standardnature]{sn-jnl}
\jyear{2022}%

\usepackage[version=4]{mhchem}
\theoremstyle{thmstyleone}%
\theoremstyle{thmstyletwo}%

\theoremstyle{thmstylethree}%

\usepackage{chemformula} 
\usepackage[utf8x]{inputenc} 
\usepackage[T1]{fontenc} 
\usepackage{mathrsfs}
\usepackage{braket}
\usepackage{amssymb}
\usepackage{appendix}
\usepackage{hyperref}
\usepackage{array}
\usepackage{longtable}
\usepackage{color}
\usepackage[normalem]{ulem}
\usepackage{dcolumn}
\usepackage{adjustbox}
\usepackage{tabularx}
\usepackage{tikz}
\usepackage{physics}
\usepackage{program}
\usepackage{graphicx}
\usetikzlibrary{arrows,shapes,positioning,shadows,trees}

\newcommand{\red}[1]{{\color{black} #1}}

\begin{document}

\title[ ]{Molecular Quantum Chemical Data Sets and Databases for Machine Learning Potentials}


\author*[1]{\sur{Arif Ullah}}\email{arif@ahu.edu.cn}

\author[2]{\sur{Yuxinxin Chen}}

\author*[2,3]{\sur{Pavlo O. Dral}}\email{dral@xmu.edu.cn}

\affil[1]{\orgdiv{School of Physics and Optoelectronic Engineering}, \orgname{Anhui University}, \orgaddress{\city{Hefei}, \postcode{230601}, \state{Anhui}, \country{China}}}
\affil[2]{\orgdiv{State Key Laboratory of Physical Chemistry of Solid Surfaces, College of Chemistry and Chemical Engineering, Fujian Provincial Key Laboratory of Theoretical and Computational Chemistry}, \orgname{Xiamen University}, \orgaddress{\city{Xiamen}, \postcode{361005}, \state{Fujian}, \country{China}}}
\affil[3]{Institute of Physics, Faculty of Physics, Astronomy, and Informatics, Nicolaus Copernicus University in Toruń, ul. Grudziądzka 5, 87-100 Toruń, Poland}

\abstract{
The field of computational chemistry is increasingly leveraging machine learning (ML) potentials to predict molecular properties with high accuracy and efficiency, providing a viable alternative to traditional quantum mechanical (QM) methods, which are often computationally intensive. Central to the success of ML models is the quality and comprehensiveness of the data sets on which they are trained. Quantum chemistry data sets and databases, comprising extensive information on molecular structures, energies, forces, and other properties derived from QM calculations, are crucial for developing robust and generalizable ML potentials.
In this review, we provide an overview of the current landscape of quantum chemical data sets and databases. We examine key characteristics and functionalities of prominent resources, including the types of information they store, the level of electronic structure theory employed, the diversity of chemical space covered, and the methodologies used for data creation. Additionally, an updatable resource is provided to track new data sets and databases at~\url{https://github.com/Arif-PhyChem/datasets_and_databases_4_MLPs}. \red{This resource also has the overview in a machine-readable database format with the Jupyter notebook example for analysis.}
Looking forward, we discuss the challenges associated with the rapid growth of quantum chemical data sets and databases, emphasizing the need for updatable and accessible resources to ensure the long-term utility of them. We also address the importance of data format standardization and the ongoing efforts to align with the FAIR principles to enhance data interoperability and reusability. Drawing inspiration from established materials databases, we advocate for the development of user-friendly and sustainable platforms for these data sets and databases.
}
 
\maketitle
\section{Introduction}\label{sec:introducton}

Accurate and efficient prediction of molecular properties is a cornerstone of computational chemistry. While quantum mechanical (QM) calculations provide unparalleled precision, their computational cost often limits their application to large-scale systems. To bridge this gap, machine learning (ML) potentials have emerged as promising alternatives, capable of predicting molecular properties with reasonable accuracy and remarkable efficiency.

Central to the development of robust ML potentials is the availability of high-quality training data. Quantum chemical data sets and databases serve as the foundation for these models, providing essential information on molecular structures, energies, forces, and other properties. The breadth and depth of these data sets and databases directly influence the ability of ML models to learn complex chemical phenomena and generalize to new systems. Furthermore, accessible and well-curated data sets and databases are essential for reproducibility and benchmarking in the field.

Given the plethora of ML potentials and their growing importance in computational chemistry, this review aims to provide \red{an overview} of the current landscape of \red{molecular} quantum chemical data sets and databases. We explore the key characteristics and functionalities of prominent resources, examining the type of information they store (e.g., molecular structures, various properties), the level of electronic structure theory employed (e.g., density functional theory (DFT), second-order Møller-Plesset perturbation theory (MP2)), the size and diversity of the chemical space they cover, and the methodologies used for data creation. \red{While the possible selection of the data sets an databases is very broad, we mostly focus on the data collections which were specifically designed with the goal of developing and evaluating ML potentials of molecules or have a potential to be used for such applications. Data sets for both ground and excited state potentials are included and many data sets also contain properties that go beyond energies and forces. We also included data collections that provide the sets of 3D nuclear geometries and energies which are frequently used in the context of structure-property relations such as QM9 data set\cite{ramakrishnan2014quantum} but in principle can be used to train ML potentials and, indeed, QM9 is sometimes used to benchmark ML potentials\cite{nrc20lilienfeld}. Surely, there are more data sets generated in the plenitude of studies involving MLPs, which either served to highlight the particular strengths and shortcomings of a new methodology such as a new active learning protocol or were a "by-product" of investigating an interesting chemical phenomenon. Although these data collections are undoubtedly very useful, we have to exclude them to limit the scope of our current overview. Here were we aim to focus on the data sets generated wit as little as possible involvement of ML potentials to have a collection less biased to a particular ML potential. This is admittedly hard to avoid as, e.g., ANI-1x data set\cite{smith2020ani} was generated using techniques such as normal mode sampling but the final composition was obtained via pool-based active learning with the ANI-type universal potential.

With these limitations in mind,} we introduce an updatable online resource designed to track the emerging data sets and databases, ensuring researchers have access to the latest and most relevant data (\url{https://github.com/Arif-PhyChem/datasets_and_databases_4_MLPs}). \red{This GitHub repository allows other researchers to make updates by creating pull requests or issues and it is less limited than the current overview because, e.g., the researchers can add data collections generated with active learning. In addition, for easier navigation through the data resources, we collect them in the JSON database format and provide Jupyter notebooks with examples to, e.g., convert it to the common spreadsheet formats or create the tables like Table~\ref{tab:overview} in this Review.}

This review also highlights several key challenges in the field, including ensuring the long-term accessibility of data sets, maintaining updatable resources, and advancing data format standardization in line with the FAIR principles.\cite{FAIR} By identifying these challenges, we aim to guide researchers toward the most effective use of available resources and best practices for future developments.

\section{Overview}

In Table~\ref{tab:overview}, we present a concise overview of the data sets and databases discussed here, organized in alphabetical order, as of September 2024.

\begin{longtable}{|p{2.5cm}|p{3cm}|p{3cm}|p{1.8cm}|} 
\caption{A brief overview of the data sets and databases discussed in this paper, organized alphabetically for easy reference.} \label{tab:overview} \\
\hline
\textbf{Data set / Database} & \textbf{Description} & \textbf{Methods; Software Used} & \textbf{Data Access} \\ \hline
\endfirsthead
\multicolumn{4}{c}%
{\tablename\ \thetable\ -- \textit{Continued from previous page}} \\
\hline
\textbf{Data set / Database} & \textbf{Description} & \textbf{Methods; Software Used} & \textbf{Data Access} \\ \hline
\endhead
\hline \multicolumn{4}{|r|}{\textit{Continued on next page}} \\ \hline
\endfoot
\hline
\endlastfoot

\red{AIMEl-DB~\cite{meza2024quantum}} & \red{Datasets containing atomic properties for 44K small organic molecules} & \red{B3LYP/6-31G(2df,p);Gaussian 09} & \red{\href{Zenodo}{https://zenodo.org/records/11406726}} \\ \hline

\red{Alchemy~\cite{chen2019alchemy}} & \red{19,487 organic molecules containing up to 14 heavy atoms from the GDB-17 database with 12 quantum mechanical properties} & \red{B3LYP/6-31G(2df,p), MMFF94, HF/STO-3G; PySCF, RDKit} & \red{\href{GitHub}{https://github.com/tencent-alchemy/Alchemy}} \\ \hline

ANI-1~\cite{smith2017ani} &  A collection of non-equilibrium DFT total energy calculations for organic molecules, encompassing approximately 20 million conformations of 57,462 small organic molecules & MMFF94 force field and $\omega$B97X/6-31G(d); RDKit and Gaussian 09  & \href{Figshare}{https://doi.org/10.6084/m9.figshare.5287732.v1}, \href{GitHub}{https://github.com/isayev/ANI1_dataset} \\ \hline
ANI-1x and ANI-1ccx~\cite{smith2020ani}& The ANI-1x data set includes DFT calculations for approximately 5 million organic molecule conformations while the ANI-1ccx data set is a subset of ANI-1x data set, recomputed with CCSD(T)/CBS level of theory & Bootstrapping, Four active learning sampling techniques,  $\omega$B97X/6-31G*, $\omega$B97X/def2-TZVPP, CCSD(T)*/CBS and MBIS; RDKit, ASE , Gaussian 09, ORCA, HORTON software library & \href{Figshare}{https://doi.org/10.6084/m9.figshare.c.4712477}, \href{GitHub}{https://github.com/aiqm/ANI1x_datasets} \\ \hline

\red{ANI-2x~\cite{devereux2020extending}} & \red{8.9 million nonequilibrium neutral singlet molecules with seven chemical elements (H, C, N, O, S, F, Cl)} & \red{$\omega$B97X/6-31G*; Gaussian 09} & \href{Zenodo}{https://doi.org/10.5281/zenodo.10108942} \\ \hline

bigQM7$\omega$~\cite{kayastha2022resolution} & Ground-state properties and electronic spectra for over 12,880 molecules with 7 CONF atoms, sourced from the GDB11 dataset & UFF, ConnGO workflow, 3-21G, def2SVP, def2TZVP, def2SVPD, Zindo and PM6; Gaussian & \href{Link1}{https://moldisgroup.github.io/bigQM7w}, \href{Link2}{https://dx.doi.org/10.17172/NOMAD/2021.09.30-1}, \href{Link3}{https://moldis.tifrh.res.in/index.html} \\ \hline
C7O2H10-17~\cite{schutt2017quantum}  & MD trajectories of 5,000 steps for 113 isomers of \ce{C7O2H10} & PBE functional; FHI-aims & \href{Link}{http://quantum-machine.org/datasets/} \\ \hline
CheMFi~\cite{vinod2024chemfi} & A multifidelity compilation of quantum chemical properties derived from a subset of the WS22 database, featuring 135,000 geometries sampled from nine distinct molecules. It encompasses five different levels of fidelity, each corresponding to a specific basis set size (STO-3G, 3-21G, 6-31G, def2-SVP, def2-TZVP) & TD-DFT, CAM-B3LYP, STO-3G, 3-21G, 6-31G, def2-SVP and def2-TZVP; ORCA & \href{GitHub}{https://github.com/SM4DA/CheMFi} \\ \hline

\red{COMPAS Project~\cite{wahab2022compas, mayo2024compas, wahab2024compas}} & \red{The COMPAS project provides structures and properties for polycyclic aromatic systems where COMPAS-1 (43k), COMPAS-2 (0.5 million), and COMPAS-3 (40k) focus on different types of polycyclic molecules. All contain molecules up to 11 rings (xTB), with up to 10 rings calculated at DFT level using CaGe, GFN1/2-xTB, B3LYP-D3(BJ)/def2-SVP, and CAM-B3LYP-D3BJ/aug-cc-pVDZ} & \red{GFN2-xTB, GFN1-xTB, B3LYP-D3(BJ)/def2-SVP,  CAM-B3LYP-D3BJ/def2-SVP, CAM-B3LYP-D3BJ/aug-cc-pVDZ//CAM-B3LYP-D3BJ/def2-SVP; xTB, Gage, ORCA 4.2.0} & \red{\href{GitLab}{https://gitlab.com/porannegroup/compas}} \\ \hline

\red{CREMP~\cite{grambow2024cremp}} & \red{CREMP data set (36k macrocyclic peptides, 31.3 million conformers) for ML training. Generated using CREST, RDKit, ETKDGv3, MMFF94, GFN2-xTB, and metadynamics. Properties calculated at semiempirical level. CREMP-CycPeptMPDB (8.7 million conformers) added for passive membrane permeability} & \red{GMMFF94, GFN2-xTB, ETKDGv3; CREST, xTB} & \red{\href{GitHub}{https://github.com/Genentech/cremp}, \href{Link1}{https://doi.org/10.5281/zenodo.7931444}, \href{Link-2}{https://doi.org/10.5281/zenodo.10798261}} \\ \hline

\red{GEOM~\cite{axelrod2022geom}} & \red{Conformers for mid-sized organic molecules from QM9 and experimental datasets related to physical chemistry, biophysics, and physiology} & \red{MMFF, GFN2-xTB, r2scan-3c/mTZVPP; ORCA 5.0.2, CREST, RDKit, CENSO} & \red{\href{GitHub}{https://github.com/learningmatter-mit/geom}} \\ \hline

ISO17~\cite{schutt2017schnet} & An extension of C7O2H10-17 data set, consisting of 129 isomers & PBE functional (GGA and Tkatchenko
Scheffler (TS) van der Waals correction method); FHI-aims & \href{Link}{http://quantum-machine.org/datasets/} \\ \hline

MD17 and its later versions~\cite{chmiela2017machine, christensen2020role, chmiela2018towards} & \textit{Ab initio} molecular dynamics trajectories collection along with total energy and forces for ten small organic molecules& PIMD, PBE + vdW-TS, PBE/def2-SVP, CCSD, CCSD(T) and DFT FHI-aims; i-PI code, ORCA and FHI-aims & \href{Link}{http://www.sgdml.org/#data_sets}\\ \hline

MD22~\cite{chmiela2023accurate} & MD trajectories for seven systems spanning four major classes of biomolecules and supramolecular structures & PBE+MBD level of theory with "light" and "tight" basis sets within the FHI-aims framework; FHI-aims and i-PI& \href{Link}{http://www.sgdml.org/#data_sets} \\ \hline

MultiXC-QM9~\cite{nandi2023multixc}  &  Expands upon QM9 by including data from 76 different DFT functionals alongside three basis sets and a semi-empirical method (GFN2-xTB) & DFT (76 functionals), three basis sets and GFN2-xTB; the ADF software package & \href{Figshare}{https://doi.org/10.11583/DTU.c.6185986.v3} \\ \hline

$\nabla^2$DFT~\cite{khrabrov2024nabla} & A comprehensive collection of about 16 million conformers for 2 million drug-like molecules, offering DFT-calculated energies, forces, and various properties, along with relaxation trajectories for many molecules & $\omega$B97X-D/def2-SVP; Psi4 & \href{GitHub}{https://github.com/AIRI-Institute/nablaDFT} \\ \hline

\red{OFF-ON~\cite{celerse2024offon}} & \red{Organic fragments from organocatalysts that are non-modular} & \red{DFTB3, PBE0-D3/def2-SVP; DFTB+, TeraChem} & \red{\href{Link}{https://archive.materialscloud.org/record/2023.189}} \\ \hline

OrbNet Denali~\cite{christensen2021orbnet} & A dataset used to train the OrbNet Denali model, consisting of 2.3 million geometries with energy labels calculated at the DFT and semi-empirical levels, based on the ChEMBL27 database &  GFN1-xTB, AIMD and $\omega$B97X-D3/def2-TZVP; ENTOS BREEZE, DIMORPHITE-DL and ENTOS QCORE & \href{Figshare}{https://doi.org/10.6084/m9.figshare.14883867} \\ \hline

PC9~\cite{glavatskikh2019dataset} & 99,234 distinct molecules, a subset of PubChemQC project selected based on the limitations of QM9 data set (size of up to 9 heavy atoms in the range C, N, O and F) & Same as for the PubChemQC
Database & \href{Figshare}{https://doi.org/10.6084/m9.figshare.9033977.v1}, \href{Zenodo}{https://doi.org/10.5281/zenodo.3588370}\\ \hline

PubChemQC B3LYP/6-31G*//PM6~\cite{nakata2023pubchemqc} & A collection of electronic properties for nearly 86 million molecules, encompassing a broad spectrum of essential compounds and biomolecules with molecular weights up to 1000 & B3LYP/6-31G* and PM6; GAMESS & \href{Link}{https://nakatamaho.riken.jp/pubchemqc.riken.jp/b3lyp_pm6_datasets.html} \\ \hline
PubChemQC Database~\cite{nakata2017pubchemqc} & 3 million molecules optimized for their ground states and over 2 million molecules with low-lying excited states & PM3 method, Hartree-Fock method (STO-6G basis set), B3LYP functional (VWN3) and 6-31G* basis set, TD-DFT with the B3LYP functional and 6-31+G* basis set; Open Babel, Firefly, SMASH and GAMESS & \href{Link}{https://nakatamaho.riken.jp/pubchemqc.riken.jp/b3lyp_2017.html} \\ \hline
PubChemQC PM6~\cite{nakata2020pubchemqc} & PM6 data for 221 million molecules, including optimized geometries and electronic structures & PM6 geometry optimization; Open Babel and Gaussian 09 & \href{Link}{https://nakatamaho.riken.jp/pubchemqc.riken.jp/pm6_datasets.html} \\ \hline

QCDGE~\cite{zhu2024quantum} & An extensive collection of ground and excited-state properties for 443,106 organic molecules, each containing up to ten heavy atoms, including carbon, nitrogen, oxygen, and fluorine. These molecules are sourced from well-known databases such as QM9, PubChemQC, and GDB-11. & GFN2-xTB,  DFT (B3LYP, 6-31G(d), \red{BJD3} dispersion corrections) and TD-DFT($\omega$B97X-D, 6-31 G(d)); Open Babel, xtb and Gaussian 16 package & \href{Link}{http://langroup.site/QCDGE/} \\ \hline
 
QM1B~\cite{mathiasen2024generating} & A low-resolution DFT data set, QM1B, generated using PySCF IPU, contains one billion examples of molecules with 9-11 heavy atoms. It was created from 1.09M SMILES in the GDB-11 database, computing properties like the HOMO-LUMO gap for up to 1000 conformers per molecule. & B3LYP/STO-3G; PySCF$_{\rm IPU}$ & \href{GitHub}{https://github.com/graphcore-research/qm1b-dataset} \\ \hline

QM7~\cite{rupp2012fast} & Focuses on a subset of 7,165 small organic molecules from the GDB-13 database, providing Coulomb matrices, atomization energies, atomic charge, and Cartesian coordinates &  PBE0/tier2 basis set; Open Babel and FHI-aims & \href{Link}{http://quantum-machine.org/datasets/} \\ \hline
QM7-X~\cite{hoja2021qm7} & 4.2 million structures of small organic molecules (containing up to seven non-hydrogen atoms) with a rich set of 42 properties & MMFF94, DFTB3+MBD; Open Babel, Confab, DFTB+, ASE, and FHI-aims & \href{Zenodo}{https://doi.org/10.5281/zenodo.4288677} \\ \hline

QM7b~\cite{montavon2013machine} & An extension of QM7, providing data on 7,211 small organic molecules including 14 properties such as atomization energy, static polarizability, and frontier orbital eigenvalues, calculated using DFT and other quantum chemistry methods & UFF, Zindo, SCS, PBE, PBE0, and GW calculations; Open Babel, ORCA, and FHI-aims code (tight settings/tier2 basis set) & \href{Link}{http://stacks.iop.org/NJP/15/095003/mmedia} \\ \hline
QM8~\cite{ramakrishnan2015electronic} & Provides electronic spectra data for approximately 21,786 small organic molecules, derived from QM9, calculated using TD-DFT and other excited-state methods &  CC2 method, B3LYP/6-31G(2df,p), long-range corrected TD-DFT, PBE0 and CAM-B3LYP functionals, def2SVP and def2TZVP; TURBOMOLE program & \href{Link}{https://pubs.aip.org/jcp/article-supplement/73278/zip/084111_1_supplements/} \\ \hline

QM9~\cite{ramakrishnan2014quantum} & A collection of molecular structures and properties for 134,000 small organic molecules & B3LYP/6-31G(2df,p); Gaussian 09 & \href{Figshare}{https://doi.org/10.6084/m9.figshare.978904} \\ \hline
QM9-G4MP2~\cite{kim2019energy} & Provides highly accurate G4MP2 calculations for the molecular structures within QM9 & G4MP2; Gaussian 16& \href{Figshare}{https://doi.org/10.6084/m9.figshare.c.4351631.v1} \\ \hline

QM9S~\cite{zou2023qm9s}  & A collection of 33,885 organic molecules from QM9 for training/testing DetaNet where geometries were re-optimized with Gaussian 16 (B3LYP/def-TZVP). Molecular properties (scalar, vector, tensor) were calculated at the same level, including IR, Raman, and UV-Vis spectra from frequency analysis and TD-DFT & B3LYP/def-TZVP and TD-DFT; Gaussian16 package (B.01 version) & \href{Figshare}{https://doi.org/10.6084/m9.figshare.24235333} \\ \hline

QM-sym~\cite{liang2019qm} & 135,000 structures with C$_{n\text{h}}$ symmetry & B3LYP/6-31G(2df,p); Gaussian 09 & \href{GitHub}{https://github.com/XI-Lab/QM-sym-database}, \href{Figshare}{https://doi.org/10.6084/m9.Figshare.9638093} \\ \hline
QM-symex~\cite{liang2020qm} & The QM-sym data set has been expanded to include an additional 38,000 molecules, providing valuable information on excited-state properties & TD-B3LYP/6-31G//B3LYP/6-31G(2df,p); Gaussian 09 & \href{Figshare}{https://doi.org/10.6084/m9.Figshare.12815276} \\ \hline

QM-22~\cite{bowman2022md17} & A compilation of molecular data sets specifically curated for DMC calculations of the zero-point state &  Each data set within QM22 employs unique methodologies tailored to the specific molecules involved. Detailed computational methods for each data set can be found in their corresponding publications & \href{GitHub}{https://github.com/jmbowma/QM-22} \\ \hline

QMugs~\cite{isert2022qmugs} & Quantum-mechanical properties of over 665,000 drug-like molecules extracted from the ChEMBL database & Merck molecular force field, GFN2-xTB and $\omega$B97X-D/def2-SVP; RDKit, xTB software package and Psi4 & \href{Link}{https://doi.org/10.3929/ethz-b-000482129} \\ \hline

\red{revQM9~\cite{khan2024revqm9}} & \red{The revision of QM9 data set with the quantum chemical properties recalculated at the aPBE0 level} & \red{aPBE0/cc-pVTZ; PySCF 2.4.0} & \red{\href{Zenodo}{https://zenodo.org/doi/10.5281/zenodo.10689883}} \\ \hline

SPICE~\cite{eastman2023spice, eastman2024spice} & 2,008,628 conformations of 113,999 drug-like small molecules and proteins & $\omega$B97M-D3BJ/def2-TZVPPD; Psi4 & \href{Zenodo}{https://doi.org/10.5281/zenodo.7338495}, \href{GitHub}{https://github.com/openmm/spice-dataset} \\ \hline

TensorMol ChemSpider~\cite{yao2018tensormol} & 3 million conformations from 15k different molecules, sourced from the ChemSpider database  & $\omega$B97X-D and 6-311G**; QChem program & \href{Link}{https://drive.google.com/drive/folders/1IfWPs7i5kfmErIRyuhGv95dSVtNFo0e_} \\ \hline

\red{tmQM~\cite{balcells2020tmqm}} & \red{86k transition metalorganic molecules based on Cambridge structural database} & \red{GFN2-xTB, TPSSh/def2-SVP; xTB, Gaussian 16} & \red{\href{GitHub}{https://github.com/bbskjelstad/tmqm}, \href{Link}{http://quantum-machine.org/datasets/}} \\ \hline

Transition1x~\cite{schreiner2022transition1x}  &  9.6 million data points, each meticulously generated using DFT calculations with forces and energies for a staggering 10,000 organic reactions & $\omega$B97X, 6-31G(d), NEB and CINEB; ASE and ORCA  & \href{Figshare}{https://doi.org/10.6084/m9.figshare.19614657.v4}, \href{GitLab}{https://gitlab.com/matschreiner/T1x} \\ \hline

VIB5~\cite{zhang2022vib5} & Ab initio quantum chemical data for five small polyatomic molecules with significant astrophysical relevance & MP2/cc-pVTZ, CCSD(T)/cc-pVQZ and HF with cc-pVTZ and cc-pVQZ basis sets, CCSD(T) with extrapolated complete basis set, core-valence electron correlation energy corrections, and diagonal Born--Oppenheimer corrections; CFOUR and MOLPRO & \href{Figshare}{https://doi.org/10.6084/m9.figshare.16903288} \\ \hline

VQM24~\cite{khan2024towards} & Providing QM properties of 258,242 unique constitutional isomers and 577,705 conformers of varying stoichiometries, focusing on molecules composed of up to five heavy atoms from C, N, O, F, Si, P, S, Cl, Br &  MMFF94, GFN2-xTB, $\omega$B97X-D3/cc-pVDZ and DMC@PBE0/ccECP/ccpVQZ; Surge, RDKit, Crest, Psi4 and QMCPACK &  \href{Zenodo}{https://doi.org/10.5281/zenodo.11164951} \\ \hline

WS22~\cite{pinheiro2023ws22} Database & 1.18 million molecular geometries (equilibrium and non-equilibrium) for ten organic molecules containing up to 22 atoms & Wigner sampling approach, geometry interpolation scheme, PBE0/6-311G*; Newton-X$\cdot$CS and Gaussian 09 & \href{Zenodo}{https://doi.org/10.5281/zenodo.7032334} \\ \hline

xxMD~\cite{pengmei2024beyond} & Excited-state molecular dynamics  data set for four molecular systems chosen for their photochemical activity: azobenzene, malonaldehyde, stilbene, and dithiophene & Surface hopping dynamics, SA-CASSCF electronic theory, unrestricted KS-DFT (M06 and 6-31G); SHARC, OpenMolcas 22.06 and Psi4& \href{GitHub}{https://github.com/zpengmei/xxMD}, \href{Zenodo}{https://doi.org/10.5281/zenodo.10393859} \\ \hline
\end{longtable}

\section{Data Sets and Databases}

\subsection{QM9}
 QM9 data set is one of the widely used data sets, a collection of molecular structures and properties for 134,000 small organic molecules.\cite{ramakrishnan2014quantum} QM9 targets neutral organic molecules containing up to nine non-hydrogen atoms (C, N, O, and F). This selection corresponds to the GDB-9 subset of the larger GDB-17 database.\cite{ruddigkeit2012enumeration}  While ensuring a manageable size, it captures a significant portion of organic chemical space. Notably, the data set incorporates relevant biomolecules like amino acids (glycine, alanine) and nucleobases (cytosine, uracil, thymine), alongside pharmaceutically important building blocks (pyruvic acid, piperazine, hydroxyurea). Among the 134,000 molecules, 621 distinct chemical formulas (stoichiometries) are present, with \ce{C7H10O2} being the most abundant, containing 6,095 constitutional isomers.

\vspace{0.4cm}

\noindent \textit{Computational Methodology}:
  QM9 focuses on organic molecules with selection process opting for neutral molecules and excluding cations and anions. While, zwitterions are kept due to their importance in biomolecules like amino acids, heavier halogens (sulfur, bromine, chlorine, iodine) are excluded from the selection. Initial 3D structures were generated using SMILES strings followed by geometry optimization using computational methods (pre-optimization at PM7 semi-empirical level with MOPAC software\cite{stewart1990mopac} and final optimization at B3LYP/6-31G(2df,p)\cite{beck1993density, stephens1994ab, ditchfield1971a, krishnan1980a} with the Gaussian 09 software\cite{frisch2009gaussian}). Stringent convergence criteria were employed to guarantee high-quality structures. Following geometry optimization, the B3LYP/6-31G(2df,p) level was employed to calculate various properties for each molecule. Notably, for 6,095 constitutional isomers of \ce{C7H10O2}, a more accurate method G4MP2\cite{curtiss2007gaussian} was employed for energetics calculations.  
 
\vspace{0.4cm}

\noindent \textit{Data Accessibility}:
QM9 data set is publicly available on Figshare at \url{https://doi.org/10.6084/m9.figshare.978904}.\cite{ramakrishnan2014fig} For each molecule, atomic coordinates and calculated properties are stored in a file named \texttt{dataset\_index.xyz}. The included properties are equilibrium geometries, atomization energies, enthalpies, entropies, frontier orbital eigenvalues, dipole moments, harmonic frequencies, polarizabilities, and thermochemical energetics.  

\subsection{QM9-G4MP2}
The QM9-G4MP2 database\cite{kim2019energy} is built upon the existing QM9 data set by providing highly accurate G4MP2 (Gaussian-4 theory using reduced order perturbation\cite{curtiss2007gaussian, curtiss2007gtheory}) calculations for the molecular structures within QM9. The G4MP2 approach employed in this database comprises the calculations with a series of methods including DFT, Hartree--Fock (HF), Møller--Plesset perturbation theory (MP2), and coupled-cluster single and double excitations with perturbative triple correction (CCSD(T)). 

\vspace{0.4cm}

\noindent \textit{Computational Methodology}:
To generate the QM9-G4MP2 data set, the geometries of all QM9 molecules underwent reoptimization using the B3LYP/6-31G(2df,p) level theory, followed by single-point energy calculations employing CCSD(T,FC)/6-31G(d), MP2(FC)/G3MP2largeXP, RHF/mod-aug-cc-pVTZ, and RHF/mod-aug-cc-pVQZ methods. These calculations were executed using the Gaussian~16 package (version A.03)
\cite{g16} utilized in the original QM9. Additionally, to derive accurate atomization energies, G4MP2 calculations were performed for individual atoms of H, C, N, O, and F, maintaining identical charge and spin multiplicity as those within the QM9 molecules. Note that these calculations were conducted in varying computational environments and with Gaussian 16 instead of Gaussian 09 employed in QM9.

\vspace{0.4cm}

\noindent \textit{Data Accessibility}:
The QM9-G4MP2 database is accessible in a compressed file format via Figshare at \url{https://doi.org/10.6084/m9.figshare.c.4351631.v1}.\cite{kim2019fig} The database includes all raw calculation outputs, scripts utilized for database construction, and pre-processed data presented in comma-separated values (CSV) format.

\subsection{MultiXC-QM9}
The MultiXC-QM9 data set\cite{nandi2023multixc} significantly expands upon the well-established QM9 data set for studying small molecules. While QM9 focuses on a single functional (B3LYP) and basis set, MultiXC-QM9 offers a richer resource, encompassing 76 different DFT functionals alongside three basis sets and a complementary semi-empirical method (GFN2-xTB).\cite{bannwarth2021extended} Beyond molecular energies, MultiXC-QM9 provides energies for all possible monomolecular interconversions (A$\leftrightarrow$B type) within the QM9 data set, calculated using the same methodology as the molecular energies. Additionally, the data set includes information on bond changes associated with these 162 million reactions.

This multifaceted nature makes MultiXC-QM9 useful for delta learning, transfer learning and multitask learning. The presence of data from both high and low fidelity methods enables models to learn the relationships between these methods and improve prediction accuracy.

\noindent \textit{Computational Methodology}:
The MultiXC-QM9 data set was constructed using xyz molecular geometries obtained from the QM9-G4MP2 data set\cite{kim2019fig}. Molecular energies were computed using the ADF package,\cite{te2001chemistry} which implements various post-self-consistent field (SCF) methods for different functionals. Specifically, the PBE functional\cite{perdew1996generalized} was employed for generalized gradient approximation (GGA) level energy calculations, with SZ, DZP, and TZP basis sets; these calculations were performed for 133k molecules. Additionally, single-point energies were computed using the GFN2-xTB semi-empirical method. 

For the DFT calculations, the default SCF convergence criteria of the ADF software package were utilized, set to $1\cdot10^{-6}$ hartree. Similarly, the SCF convergence for the GFN2-xTB calculations was set to the default value of $1\cdot10^{-6}$ hartree.

In the calculation of reaction energies, all possible "A$\leftrightarrow$B type" isodesmic reactions were identified among the QM9 molecules, and their corresponding indices were saved in a separate CSV file named "reactions.csv". Subsequently, the energy changes associated with each reaction were computed using the same level of theory as for the energy calculations of individual molecules and stored in additional CSV files.

\vspace{0.4cm}

\noindent \textit{Data Accessibility}:
The MultiXC-QM9 data set is available on Figshare platform at \url{https://doi.org/10.11583/DTU.c.6185986.v3}\cite{nandi2023fig} and provides information on both molecules and reactions in separate files. Energy calculations for molecules are provided in CSV and SQLite formats. CSV format contains energies from various DFT and semi-empirical methods, SMILES strings derived from xyz files, and chemical formulas for each molecule. The SQLite format stores xyz coordinates, energies, and other relevant properties derived automatically by atomistic simulation environment. Reaction data is provided in CSV format which includes reaction energies, indices of reactants and products, and SMILES strings for both reactants and products.

\red{

\subsection{revQM9}
The revQM9 data set~\cite{khan2024revqm9} is the revision of the QM9 data set with the quantum chemical properties recalculated at the aPBE0 level (which is PBE0 method reformulation with the ratio of exact exchange determined by an ML model to approximate the CCSD(T)-level energies).

\vspace{0.4cm}

\noindent \textit{Computational Methodology}:
The data set is the revision of the QM9 data set with the properties of ca. 130k equilibrium geometries (with up to 9 non-hydrogen atoms) calculated at a more accurate aPBE0 level with the cc-pVTZ basis set. PySCF 2.4.0\cite{pyscf1,pyscf2,pyscf3} was used to perform simulations, while the adaptive parameters were supplied to it by a machine learning model trained on so that aPBE targets the CCSD(T)/cc-pVTZ atomization energies. The code to make predictions with aPBE0 is available on GitHub.\cite{apbe0} The properties calculated at aPBE/cc-pVTZ include total and atomization energies, orbital energies, dipole moments, and density matrices.

\vspace{0.4cm}

\noindent \textit{Data Accessibility}:
The revQM9 data set is available at \url{https://zenodo.org/doi/10.5281/zenodo.10689883}.\cite{revqm9zenodo} The data are provided in the npz format and sample script shows how to extract the required properties from the corresponding revQM9.npz data file.

}

\red{

\subsection{AIMEl-DB}
AIMEl-DB\cite{meza2024quantum} provides atomic properties based on electronic density for 44k molecules randomly extracted from QM9 dataset,\cite{ramakrishnan2014fig} facilitating the development of ML models for direct reactivity and spectral prediction.

\vspace{0.4cm}

\noindent \textit{Computational Methodology}:
Initial 44k structures (CHNO only) are extracted from QM9 with up to 9 atoms and single-point calculations are performed at B3LYP/6-31G(2df,p) level of theory with Gaussian~16. Atomic properties are generated using AIMAll package.\cite{keith2019aimall}

\vspace{0.4cm}

\noindent \textit{Data Accessibility}:
The dataset is publicly available at Zenodo \url{https://zenodo.org/records/11406726}. More than 30 atomic properties including energy, dipole moment, quadrupole moment and population are provided as well as molecular properties stored in Gaussian output files.\cite{aimbd2024quantum}

}

\subsection{QM7-X}
QM7-X\cite{hoja2021qm7} is a comprehensive data set of 4.2 million structures of small organic molecules containing up to seven non-hydrogen atoms (C, N, O, S, C). It includes a systematic sampling of all stable equilibrium structures (constitutional/structural isomers and stereoisomers) and explores 100 non-equilibrium structures for each molecule. Each structure in QM7-X is accompanied by a rich set of 42 properties, meticulously calculated using PBE0+MBD\cite{seifert1996calculations,tkatchenko2012accurate} level theory. These properties range from fundamental aspects like atomization energies and dipole moments to response characteristics like polarizability and dispersion coefficients. 
\vspace{0.4cm}

\noindent \textit{Computational Methodology}: To generate QM7-X data set, researchers first used the GDB13 database\cite{blum2009970} to identify all possible structures (constitutional/structural isomers and stereoisomers) for molecules with up to seven non-hydrogen atoms. Using SMILES strings, initial 3D structures were obtained using MMFF94 force field\cite{halgren1996merck} followed by conformational isomer search using Confab\cite{o2011confab} to generate various conformers for each molecule. All structures were re-optimized at DFTB3+MBD level theory.\cite{seifert1996calculations,tkatchenko2012accurate} The lowest-energy structure was selected as the first conformer, and additional structures with RMSD (root mean square deviation) larger than 1.0 {\AA} from existing conformers were included.

In the case of non-equilibrium structures, 100 non-equilibrium structures were created for each equilibrium structure. It was accomplished by displacing the equilibrium structures along linear combinations of normal mode coordinates computed at the DFTB3+MBD level within the harmonic approximation. It was ensured that the generated structures follow a Boltzmann distribution with a specific average energy difference from the corresponding equilibrium structure.

This process resulted in about 4.2 million equilibrium and non-equilibrium structures forming the QM7-X data set. Physicochemical properties were calculated on the generated structures using PBE0+MBD level of theory with the FHI-aims code.\cite{blum2009ab} Tight settings were used for basis functions and integration grids. Various properties were calculated including energies, forces, atomic charges, dipole moments, polarizabilities, and more.

\vspace{0.4cm}

\noindent \textit{Data Accessibility}:
The QM7-X data set is available in eight HDF5 files on Zenodo at \url{https://doi.org/10.5281/zenodo.4288677}.\cite{hoja2021zenodo} Each file stores information about molecular structures using Python dictionaries. These dictionaries contain various properties for each molecule like atomic numbers, positions (coordinates), and other physicochemical properties.

\subsection{QM7}
Curated from the much larger GDB-13 database\cite{blum2009970} containing nearly 1 billion molecules, QM7 focuses on a subset of 7,165 small organic molecules.\cite{rupp2012fast}  These molecules contain up to 23 atoms, with a maximum of 7 being "heavy atoms" -C, N, O, and S. The remaining positions are filled with hydrogen atoms. The data set provides Coulumb matrices, atomization energies, atomic charge and the Cartesian coordinate of each atom in the molecules.
\vspace{0.4cm}

\noindent \textit{Computational Methodology}:
The researchers selected a subset of 7,165 molecules from GDB-13 with atomization energies ranging from $-800$ to $-2000$~kcal/mol. This subset encompasses molecules with features like double and triple bonds, cycles, and various functional groups including carboxy, cyanide, amide, alcohol, and epoxy groups. In addition, only constitutional isomers (molecules with different chemical bond arrangements) were included and conformational isomers were excluded. For calculations, researchers first used Open Babel software\cite{guha2006blue} to generate 3D structures for each molecule from the subset. Next, they employed DFT calculations with PBE functional/tier2 basis set implemented in the FHI-aims code to calculate the atomization energy for each molecule with high accuracy.
\vspace{0.4cm}

\noindent \textit{Data Accessibility}:
 The entire data set is hosted publicly on Quantum-Machine.org\cite{rupp2012qm7} with data organized into several multidimensional arrays. It provides Coulomb matrices, atomization energies, atomic charge and the Cartesian coordinate of each atom in the molecules.

\subsection{QM7b}
The QM7b data set\cite{montavon2013machine} is a collection of information on over 7211 small organic molecules containing up to seven different elements: C, Cl, H, N, O, and S. It serves as an extension of the QM7 data set specifically designed for multitask learning applications in chemistry. Each molecule is described by 14 properties, including atomization energy, static polarizability, frontier orbital eigenvalues (HOMO and LUMO) and excitation energies.

\vspace{0.4cm}

\noindent \textit{Computational Methodology}:
The creation of QM7b data set starts from the selection of 7211 small organic molecules from GDB-13 database. Initial molecular geometries were generated from SMILES strings\cite{weininger1988smiles} using the Universal Force Field (UFF)\cite{rappe1992uff} within the Open Babel software. Subsequent geometry optimization was performed using the PBE approximation to Kohn--Sham DFT within the FHI-aims code. For molecular electronic properties, the PBE0 DFT\cite{seifert1996calculations} was employed to calculate atomization energies and frontier orbital eigenvalues for each molecule. The ZINDO approach\cite{stewart1990mopac} was used to determine properties like electron affinity, ionization potential, excitation energies, and maximal absorption intensity. The GW approximation\cite{hedin1965new} was utilized to evaluate the frontier orbital eigenvalues. For static polarizability calculations, both self-consistent screening (SCS)\cite{tkatchenko2012accurate} and PBE0 were used. Software-wise, FHI-aims was used for SCS, PBE0, and GW calculations, while ORCA code\cite{neese2008ab} handled ZINDO/s calculations. 

\vspace{0.4cm}

\noindent \textit{Data Accessibility}:
The complete QM7b data set is available as supplementary material\cite{montavon2012qm7b} in Ref.~\citenum{montavon2013machine}. It consists of two key files: coulomb.txt which contains unsorted Coulomb matrices (in atomic units) for all 7211 molecules while properties.txt stores the 14 properties associated with each molecule. 

\subsection{QM8}
The QM8 data set, introduced by Ramakrishnan et al.\cite{ramakrishnan2015electronic}, comprises electronic spectra data for approximately 21,786 small organic molecules. It is a carefully curated subset derived from the larger QM9 data set, which includes 134,000 molecules. QM8 offers information on electronic spectra, specifically at the level of time-dependent DFT(TD-DFT) and second-order approximate Coupled Cluster (CC2), presenting the lowest two singlet transition energies and their corresponding oscillator strengths.

\vspace{0.4cm}

\noindent \textit{Computational Methodology}:
QM8 originates from the QM9 data set, containing 134,000 molecules with up to nine heavy atoms. Initially, molecules with high steric strain in their initial geometries, computed using the B3LYP/6-31G(2df,p) method, were eliminated. Subsequently, the data set was narrowed down to molecules containing a maximum of eight heavy atoms (carbon, nitrogen, oxygen, and sulfur), resulting in approximately 21,800 molecules. The geometries for these molecules were assumed to be relaxed, as per the original QM9 data set. Single-point calculations were performed with the TURBOMOLE program\cite{furche2014turbomole} to determine ground ($S_0$) and the lowest two singlet excited states ($S_1$ and $S_2$). The used methods include long-range corrected TD-DFT (LR-TDDFT)\cite{furche2002adiabatic} with PBE0 and CAM-B3LYP functionals,\cite{perdew1996rationale} and different basis sets (def2-SVP and def2-TZVP),\cite{weigend2005balanced} as well as the resolution-of-identity approximate coupled cluster with singles and doubles substitution (RI-CC2) method\cite{hattig2000cc2} with the def2-TZVP basis set.\cite{weigend2005balanced}  A total of 14 "exotic" molecules were excluded due to convergence issues (7 molecules) or negative lowest transition energy, potentially attributed to orbital relaxation.

\vspace{0.4cm}

\noindent \textit{Data Accessibility}:
The QM8 data set is accessible as a supplementary text file in Ref.~\citenum{ramakrishnan2015electronic}. It provides indices for all molecules, facilitating the retrieval of their geometries from the QM9 data set, along with TD-DFT and CC2 excitation energies.

\red{
\subsection{Alchemy}
The Alchemy data set\cite{chen2019alchemy} contains 12 quantum mechanical properties for 119,487 organic molecules with up to 14 heavy atoms, sampled from the GDB-17 database.\cite{ruddigkeit2012enumeration} Quantum mechanical calculations were performed using DFT (B3LYP/6-31G(2df,p)) via PySCF, providing data on molecular geometries, electronic, and thermochemical properties.

\vspace{0.4cm}

\noindent \textit{Computational Methodology}:
The generation of molecular data in Alchemy was carried out using the PySCF. The quantum mechanical properties were calculated using DFT with the B3LYP functional and the 6-31G(2df,p) basis set. These calculations encompassed ground state equilibrium geometries, electronic properties, and thermochemical properties.

To ensure the reliability of molecular geometries, a structured three-stage optimization process was employed. Initial geometries were generated from SMILES strings using the MMFF94 force field, followed by HF optimization with the STO-3G basis set. The final relaxation step utilized B3LYP/6-31G(2df,p) to achieve accurate equilibrium geometries. Additional methods, such as meta-Lowdin population analysis, were employed to compute atomic charges, ensuring the transferability of these values across different molecular environments.

In total, the data generation required about 3,000,000~CPU hours, reflecting the computational complexity involved in calculating the properties of nearly 120,000 molecules. The dataset is provided in the SD file format to include detailed bond information.

\vspace{0.4cm}

\noindent \textit{Data Accessibility}:
The Alchemy data set is available at \url{https://github.com/tencent-alchemy/Alchemy}.\cite{alchemy2019github}
}

\subsection{QM1B}
QM1B, introduced by Mathiasen et al.\cite{mathiasen2024generating}, is a colossal data set comprising one billion training examples tailored for ML applications in quantum chemistry. It encompasses molecules featuring 9--11 heavy atoms and includes properties such as energy and HOMO--LUMO gap.
\vspace{0.4cm}

\noindent \textit{Computational Methodology}:
The creation of the QM1B data set commenced with 1.09 million SMILES strings sourced from the GDB-11 database,\cite{fink2005virtual, fink2007virtual} focusing on molecules with 9--11 heavy atoms. Subsequently, hydrogen atoms were added to these molecules using RDKit.\cite{rdkitsoft} Each molecule underwent the generation of up to 1000 conformers employing the ETKDG algorithm within RDKit. This process yielded a total of 305.8 million, 568.7 million, and 205.4 million conformers for molecules with 9, 10, and 11 heavy atoms, respectively. Utilizing PySCF$_\text{IPU}$, HOMO--LUMO energies were computed for the resulting one billion conformers. Notably, a trade-off was made between data set size and data quality (DFT accuracy) to achieve the extensive data set size. Consequently, the DFT calculations (B3LYP/STO-3G) employed in QM1B may exhibit less accuracy compared to other data sets such as QM9 and PC9.

\vspace{0.4cm}

\noindent \textit{Data Accessibility}:
The QM1B data set is accessible on the GitHub platform at \url{https://github.com/graphcore-research/qm1b-dataset}.\cite{mathiasen2023git} However, it is important to highlight that QM1B is released solely for research purposes, and caution is advised for applications necessitating high accuracy. The creators encourage further exploration into the implications of reduced DFT accuracy in downstream tasks.

\subsection{SPICE}
The SPICE data set\cite{eastman2023spice, eastman2024spice}, an acronym for Small-molecule/Protein Interaction Chemical Energies, focuses on the energetic interplay between drug-like small molecules and proteins. It is an essential resource for training models that can accurately predict forces and energies across a diverse array of molecules and conformations commonly encountered in drug discovery simulations.

The initial version of SPICE featured over 1.1 million molecular conformations, covering a wide chemical space that included drug molecules, dipeptides, and solvated amino acids. It encompassed 15 different elements, spanned both charged and uncharged states, and included a range of high and low energy conformations, capturing a broad spectrum of covalent and non-covalent interactions.

The updated version\cite{eastman2024spice} enhances this data set by adding over 20,000 new molecules, improving the sampling of non-covalent interactions, and including a new subset of PubChem molecules containing boron and silicon, which were not present in the initial version. Additionally, several calculations from version 1 have been corrected and incorporated into this latest release.

\noindent \textit{Computational Methodology}:
In the computation of energies and gradients (forces) for each conformation within the SPICE data set, Psi4\cite{smith2020psi4} serves as the primary tool. The calculations employ DFT with the $\omega$B97M-D3(BJ) functional\cite{najibi2018nonlocal} and the def2-TZVPPD basis set\cite{weigend2005balanced}. Beyond forces and energies, SPICE also encompasses additional properties such as dipole and quadrupole moments, atomic charges, and bond orders.

\vspace{0.4cm}

\noindent \textit{Data Accessibility}:
The SPICE data set is readily accessible on Zenodo at \url{https://doi.org/10.5281/zenodo.7338495}, stored within a single HDF5 file named SPICE-1.1.2.hdf5.\cite{eastman2022zenodo} Additional details and scripts can be found on GitHub at \url{https://github.com/openmm/spice-dataset}.

\subsection{PubChemQC}
The PubChemQC database\cite{nakata2017pubchemqc} is a valuable asset for computational chemistry research, especially in material development, drug design, and ML applications. It contains electronic structures for a vast number of molecules, including approximately 3 million molecules optimized using DFT at the B3LYP/6-31G* level in ground states and over 2 million molecules with low-lying excited states calculated via TD-DFT at B3LYP/6-31+G*.

\vspace{0.4cm}

\noindent \textit{Computational Methodology}:
The construction of the PubChemQC database involved several essential steps. Firstly, data acquisition was carried out by downloading public structure data files (SDFs) from the PubChem Project FTP site,\cite{kim2023pubchem} each containing information like InChI, SMILES representations, and molecular weight for around 25,000 molecules. Data preprocessing followed, retaining only relevant data (CID, InChI, and weight) for the database and excluding molecules with missing CIDs, unsuitable structures (molecules with $\eta^5$ bonds, ionic salts, water mixtures etc.) or isotopic variations. All molecules were considered neutral.

The next phase involved geometry optimization. Initial 3D structures were generated from InChI data using Open Babel and underwent a series of optimizations. This included an initial optimization with the PM3 method, further optimization with the HF method (STO-6G basis set), and final optimization with the B3LYP functional (VWN3) and 6-31G* basis set. Subsequently, excited state calculations were performed using optimized geometries and Time-Dependent DFT (TDDFT) with the B3LYP functional and 6-31+G* basis set.

\vspace{0.4cm}

\noindent \textit{Data Accessibility}:
Regarding data accessibility, the PubChemQC database was previously available at \url{https://pubchemqc.riken.jp} as reported in Ref.~\citenum{nakata2017pubchemqc}. However, it has been temporarily moved to \url{https://nakatamaho.riken.jp/pubchemqc.riken.jp/b3lyp_2017.html} for continued access.\cite{nakata2017pub}

\subsection{PubChemQC PM6}

The PubChemQC PM6 data sets,\cite{nakata2020pubchemqc} stands as one of the most extensive compilations in its domain, encompassing PM6 data for a staggering 221 million molecules. These data sets are based on PubChem Compounds database,\cite{kim2023pubchem} and provide optimized geometries, electronic structures, and other pivotal molecular properties. The data sets incorporate not only neutral states of molecules but also consider cationic, anionic, and spin-flipped states. 

\vspace{0.4cm}

\noindent \textit{Computational Methodology}:
The construction of the PubChemQC PM6 data sets commenced with the comprehensive retrieval and parsing of the entire PubChem Compound database. Subsequently, key molecule attributes such as weight, InChI, SMILES, and formula were extracted, with exclusion criteria set for molecules exceeding 1000 g/mol in weight. Leveraging Open Babel, initial 3D structures were generated from SMILES encodings. Gaussian 09 software facilitated PM6 geometry optimization for each molecule, with subsequent calculations extending to cationic, anionic, and spin-flipped states. Rigorous validation ensured the fidelity of optimized InChI compared to the original counterparts. 

\vspace{0.4cm} 

\noindent \textit{Data Accessibility}:
The PubChemQC PM6 data sets, previously accessible at \url{http://pubchemqc.riken.jp/pm6_datasets.html} as referenced in Ref.~\citenum{nakata2020pubchemqc}, are now hosted at \url{https://nakatamaho.riken.jp/pubchemqc.riken.jp/pm6_datasets.html} for ongoing accessibility.\cite{nakata2020pub}

\subsection{PubChemQC B3LYP/6-31G*//PM6}
The PubChemQC B3LYP/6-31G*//PM6 data set\cite{nakata2023pubchemqc} is a massive resource for researchers in chemistry, materials science, and drug discovery. It offers electronic properties for an astonishing 85,938,443 molecules (nearly 86 million), encompassing a broad spectrum of essential compounds and biomolecules with molecular weights up to 1000. This data set represents a significant portion (94\%) of the PubChem Compound catalog as of August 29, 2016. For researchers with specific needs, the data is further divided into five sub-collections. These subsets focus on molecules containing particular elements (like CHON) and have molecular weight limitations (under 300 or 500). 

\vspace{0.4cm}
\noindent \textit{Computational Methodology}: 
As a first step in the workflow of data generation, molecules were extracted from the PubChemQC PM6 database,\cite{nakata2020pubchemqc} ensuring a diverse set of known molecules. Open Babel software was used to generate input files compatible with the GAMESS quantum chemistry program for the target molecules. For most elements, electronic properties were computed using B3LYP functional with the 6-31G* basis set. For heavier elements, specific basis sets were employed, and effective core potentials were applied for certain metals. These calculations yielded various properties, including orbital energies, dipole moments, and more.

\vspace{0.4cm}
\noindent \textit{Data Accessibility}:
The PubChemQC B3LYP/6-31G*//PM6 data set is freely available for download at \url{https://nakatamaho.riken.jp/pubchemqc.riken.jp/b3lyp_pm6_datasets.html} under a Creative Commons license.\cite{nakata2023pub} Researchers can access the data in three formats: GAMESS program input/output files, selected JSON output files and PostgreSQL database. This diverse range of formats allows researchers to choose the option that best suits their specific needs and computational tools.

\subsection{PC9}
The PC9 data set\cite{glavatskikh2019dataset} serves as a valuable counterpart to the widely used QM9 data set. Unlike QM9's purely theoretical approach, PC9 leverages the PubChemQC project,\cite{nakata2017pubchemqc} a vast database of 3 million real-world molecules with calculated properties at the B3LYP/6-31G(d) level of theory.
\vspace{0.4cm}

\noindent \textit{Computational Methodology}:
To ensure compatibility with QM9, PC9 is restricted to molecules containing a maximum of 9 "heavy atoms" (excluding H) from the elements H, C, N, O, and F. After imposing the heavy atom limit, PC9 undergoes a process to eliminate duplicates stemming from factors like enantiomers, tautomers, isotopes, and potential artifacts within the PubChem database, which could result in identical-looking molecules. Consequently, PC9 ends up with 99,234 distinct molecules. This subset is divided into two groups: one consisting of molecules shared with QM9 (18,357 compounds) and the other comprising molecules unique to PubChemQC (80,877 compounds). Unlike QM9, which is confined to closed-shell neutral compounds, PC9 includes molecules with multiplicities exceeding 1.
Furthermore, due to variations in calculation methods, PC9 only provides the total energies without zero-point vibrational energies, unlike QM9 which provides much broader coverage of different energies.

\vspace{0.4cm}

\noindent \textit{Data Accessibility}:
The PC9 data set is readily accessible on Figshare and Zenodo.\cite{glavatskikh2019fig,glavatskikh2019zenodo} where it provides ground state and orbital properties at the B3LYP/6-31G* level of theory.

\subsection{bigQM7$\omega$}
The bigQM7$\omega$\cite{kayastha2022resolution} data set is a valuable resource for researchers focused on developing ML models to predict electronic spectra of molecules. This data set includes ground-state properties and electronic spectra for over 12,880 molecules, offering a broader range of structures compared to previous data sets like QM7\cite{blum2009970} and QM9.\cite{ramakrishnan2014quantum} Notably, bigQM7$\omega$ emphasizes electronic excitations and provides data across various theoretical levels. 

\vspace{0.4cm}
\noindent \textit{Computational Methodology}: 
The creation of bigQM7$\omega$ involved a detailed, multi-step process. Initially, SMILES strings for all molecules were extracted from the GDB-11 data set.\cite{fink2005virtual, fink2007virtual} These SMILES strings were then converted into initial structures using the UFF.\cite{rappe1992uff} Geometry optimization was performed using a three-tier connectivity preserving geometry optimizations (ConnGO) workflow,\cite{senthil2021troubleshooting} ensuring structurally sound geometries and mitigating rearrangement issues during high-throughput calculations. Tight convergence criteria and basis sets def2-SVP and def2-TZVP\cite{weigend2005balanced} were employed for $\omega$B97X-D\cite{chai2008long} DFT optimizations. For excited state calculations, various methods were used: ZINDO calculations were performed at PM6 minimum energy geometries, and TDDFT calculations were carried out with 3-21G, def2-SVP, def2-TZVP, and def2-SVPD basis sets. Harmonic frequency analysis confirmed local minima for optimized geometries. Three molecules failing the ConnGO connectivity test were excluded due to a substructure prone to dissociation. All calculations were executed using the Gaussian suite of programs.

\vspace{0.4cm}
\noindent \textit{Data Accessibility}:
The bigQM7$\omega$ data set offers multiple access points to facilitate exploration and utilization by researchers. The core data, encompassing structures, ground state properties, and electronic spectra, is conveniently available for download at \url{https://moldisgroup.github.io/bigQM7w}. For deeper analysis, the NOMAD repository at \url{https://dx.doi.org/10.17172/NOMAD/2021.09.30-1} provides the input and output files from the corresponding calculations. Additionally, a data-mining platform is available at \url{https://moldis.tifrh.res.in/index.html}.\cite{kayastha2022bigQM7}

\subsection{QMug}
The QMugs (Quantum-Mechanical Properties of Drug-like Molecules) data set\cite{isert2022qmugs} provides a rich resource for researchers in drug discovery and computational chemistry.  This collection curates over 665,000 molecules relevant to biology and pharmacology, extracted from the ChEMBL database.\cite{mendez2019chembl} Notably, QMugs incorporates QM properties computed using a combination of methods, specifically leveraging both semi-empirical and DFT approaches.

\vspace{0.4cm}
\noindent \textit{Computational Methodology}: 
The construction of the QMugs data set involved a multi-step process designed to capture a comprehensive picture of drug-like molecules at both the structural and QM level. The first step focused on data extraction and SMILES processing where molecules were selected from ChEMBL database\cite{mendez2019chembl} based on specific criteria, including having well-defined single-protein targets and associated activity information. This ensures the molecules hold potential relevance for drug discovery. Following extraction, the data underwent a series of cleaning steps. Molecules were neutralized to remove charged states, and extraneous components like salts and solvents were eliminated. Additionally, outlier removal techniques filtered out fragments, molecules with extreme atom counts (outside the range of 3-100 heavy atoms), and those with unaddressed radical species or persistent net charges. 

The second step addressed conformer generation and optimization where researchers employed a semi-empirical method called GFN2-xTB\cite{grimme2017robust, bannwarth2019gfn2} to generate three conformations for each drug-like molecule. Further refinement was achieved through minimization using a force field and meta-dynamics simulations. Finally, the conformations were clustered, and the lowest-energy structure from each cluster was chosen for further calculations. This approach ensured the inclusion of representative conformers for each molecule while maintaining computational efficiency.

The final step involved higher-level quantum chemical calculations. The optimized geometries obtained in the previous step served as the basis for single-point electronic structure calculations. Here, the researchers employed DFT with the $\omega$B97X-D functional and def2-SVP basis set\cite{weigend2005balanced} calculating a wider range of properties compared to the semi-empirical method used for conformer generation. The calculated properties include formation energies, dipole moments, and crucially, wave functions.

\vspace{0.4cm}
\noindent \textit{Data Accessibility}:
The data set is hosted on the ETH Library Collection service and can be downloaded at \url{https://doi.org/10.3929/ethz-b-000482129}.\cite{isert2022dataset} The data is provided in multiple formats: (1) A summary.csv file containing computed molecular properties and annotations; (2) compressed tarball files containing molecule structures (SDFs) with embedded properties, grouped by ChEMBL identifiers and (3) separate compressed tarball files for vibrational spectra and wave function data.

\subsection{OrbNet Denali}
The OrbNet Denali dataset\cite{christensen2021orbnet} is a comprehensive training collection used to develop OrbNet Denali, a ML-enhanced semiempirical method for electronic structure calculations. It includes over 2.3 million molecular geometries with corresponding energy labels calculated at the DFT and semi-empirical levels. Based on ChEMBL27 database,\cite{zdrazil2024chembl} it covers a wide range of organic molecules, including various protonation states, tautomers, non-covalent interactions, and common salts. The dataset features key elements (H, Li, B, C, N, O, F, Na, Mg, Si, P, S, Cl, K, Ca, Br, I) essential to biological and organic systems.

\vspace{0.4cm}
\noindent \textit{Computational Methodology}:
The OrbNet Denali data set is meticulously constructed to provide a broad chemical landscape relevant to both biological and organic chemistry. Initially, the ChEMBL27 database serves as a primary source, offering a wealth of chemical structures focused on SMILES strings containing elements frequently found in biomolecules (C, O, N, F, S, Cl, Br, I, P, Si, B, Na, K, Li, Ca, Mg). To maintain manageable complexity, the data set limits molecule sizes to 50 atoms. Strings corresponding to open-shell Lewis structures or those present in a separate validation set (Hutchison conformer benchmark\cite{folmsbee2021assessing}) are excluded, resulting in a random selection of over 116 943 unique SMILES strings for neutral molecules.

To capture molecular flexibility, multiple conformations (up to four) are generated for each SMILES string using the ENTOS BREEZE conformer generator, with subsequent optimization at the GFN1-xTB level of theory\cite{grimme2017robust}. Beyond equilibrium geometries, the data set includes non-equilibrium configurations to represent the dynamic nature of molecules. Two techniques are employed: Normal-mode sampling\cite{smith2017ani}, which simulates thermal fluctuations at 300 K, and \textit{ab initio} molecular dynamics (AIMD) simulations, which mimic real-time dynamics at a higher temperature (500 K). Both methods utilize the GFN1-xTB approach, with the choice between them randomized for each molecule. This methodology yields a substantial collection of over 1 771 191 equilibrium and non-equilibrium geometries derived from ChEMBL conformers.

The data set also extends beyond neutral molecules. It includes protonation states and tautomers (over 215,000 geometries) to capture various hydrogen attachment possibilities, with contributions from the QM7b data set enriching this aspect. Additionally, salt complexes (over 271,000 geometries) are generated by pairing ChEMBL molecules with common salts, and structures from other databases (JSCH2005,\cite{jurevcka2006benchmark} BioFragment Database\cite{burns2017biofragment}) are included to represent a broader range of non-covalent interactions crucial in biological systems. Furthermore, to avoid bias towards large molecules, the data set includes a diverse collection of small molecules (over 94,000 geometries) created using common chemical motifs and further diversified by substituting atoms.

\vspace{0.4cm}
\noindent \textit{Data Accessibility}:
The OrbNet Denali training set, comprising 2.3 million geometries and corresponding energy labels, is openly accessible on FigShare at \url{https://doi.org/10.6084/m9.figshare.14883867}.\cite{denali2021figshare}

\subsection{MD17 and its later versions}
The MD17 and its later versions\cite{chmiela2017machine, christensen2020role, chmiela2018towards}  serve as foundational resources for researchers benchmarking the accuracy of force fields in molecular dynamics (MD) simulations. The two versions (the 1st version (MD17)\cite{chmiela2017machine} and the revised version (rMD17)\cite{christensen2020role}) both offer a collection of MD snapshots with DFT energies and forces for ten small organic molecules, including benzene, ethanol, malonaldehyde, uracil, toluene, salicylic acid, paracetamol, naphthalene, azobenzene, and aspirin. The 3rd version of the data set contains fewer molecules computed at the coupled cluster level of theory (CCSD(T)).\cite{chmiela2018towards} 

\vspace{0.4cm}
\noindent \textit{Computational Methodology}:
The original MD17 data set contains energy and force calculations derived from AIMD simulations of gas-phase molecules at room temperature. The electronic potential energies in this data set were obtained using Kohn--Sham DFT. However, the data set faced several limitations, such as significant numerical noise that compromised data reliability. Furthermore, the original publication lacked comprehensive details on the functional, basis set, spin-polarization, integration grid, and software used, impeding reproducibility and limiting its utility in specific chemical simulations.

To rectify these shortcomings, the revised MD17 (rMD17) data set was developed. This version includes the same ten molecules but features recalculated energies and forces using the PBE functional with the def2-SVP basis set.\cite{ernzerhof1999assessment,weigend2005balanced} The rMD17 data set also implements stringent convergence criteria and a dense integration grid, enhancing accuracy.

The coupled cluster data set was created using the same geometries as those in the DFT calculations. Energies and forces were recomputed employing all-electron CCSD(T). The Dunning’s correlation-consistent basis set cc-pVTZ was utilized for ethanol, cc-pVDZ for toluene and malonaldehyde, and CCSD/cc-pVDZ for aspirin. All calculations were executed with the Psi4 software suite.\cite{smith2020psi4}

\vspace{0.4cm}
\noindent \textit{Data Accessibility}:
All data sets are available at \url{http://www.sgdml.org/#data sets}.\cite{chmiela2017md17} The rMD17 data set can also be accessible on Figshare at \url{https://dx.doi.org/10.6084/m9.figshare.12672038}.\cite{chris2020rmd17} In addition, all data sets are provided as a class in the PYG library.\cite{fey2019fast} 

\subsection{xxMD}
Sampling in conventional MD data sets like MD17 and rMD17 is often biased towards a narrow region of the potential energy surface (PES), primarily focusing on equilibrium structures. This limited exploration restricts the ability of neural force fields (NFFs) to accurately model chemical reactions involving significant molecular deformations, such as bond breaking. To address these limitations, Pengmei et al.\cite{pengmei2024beyond} introduced the extended excited-state molecular dynamics (xxMD) data set. Similar to MD17, xxMD targets small to medium-sized gas-phase molecules but includes nonadiabatic trajectories to capture the dynamics of excited electronic states. This approach allows xxMD to encompass a broader range of nuclear configurations, including those essential for chemical reactions, such as transition states and products. Additionally, xxMD covers regions near conical intersections, which are critical for determining reaction pathways across different electronic states.

\vspace{0.4cm}
\noindent \textit{Computational Methodology}:
The xxMD data set includes four photochemically active molecular systems: azobenzene, malonaldehyde, stilbene, and dithiophene. To investigate the complex electronic structures involved in these systems, particularly in the vicinity of deformed geometries and conical intersections, nonadiabatic simulations were performed using the trajectory surface hopping algorithm implemented in the SHARC code.\cite{mai2018nonadiabatic} The state-averaged complete active space self-consistent field (SA-CASSCF) level of theory\cite{roos1980complete} was employed to accurately describe the potential energy surfaces of the multiple electronic states relevant to these systems. This subset of data, generated from SA-CASSCF calculations, is referred to as xxMD-CASSCF. The trajectories in xxMD-CASSCF exhibit energy conservation and provide potential energies and forces for the first three electronic states. All SA-CASSCF calculations were carried out using OpenMolcas 22.06.\cite{fdez2019openmolcas} 

To ensure compatibility with existing data sets (MD17 and rMD17), a new subset xxMD-DFT is created which provides recomputed ground singlet electronic state potential energies and gradients using spin-polarized KS-DFT calculations with the M06 functional\cite{zhao2008m06} for the same geometries as used in SA-CASSCF calculations. All calculations are done with the Psi4 package interfaced with ASE package.\cite{larsen2017atomic} It should be noted that the xxMD data set does not include nonadiabatic coupling vectors (NACs) for reasons described in their reference data descriptor.

\vspace{0.4cm}
\noindent \textit{Data Accessibility}:
xxMD-CASSCF and xxMD-DFT both data sets are publicly available for download on GitHub at \url{https://github.com/zpengmei/xxMD} and Zenodo at \url{https://doi.org/10.5281/zenodo.10393859}.\cite{pengmei2024dataset} The data is stored in compressed archives containing pre-split extended XYZ format files based on temporal information. These files are processed using the atomic simulation environment (ASE) software package.

\subsection{MD22}
The MD22 data set\cite{chmiela2023accurate} is a comprehensive benchmark collection featuring molecules ranging from a small peptide with 42 atoms to a double-walled nanotube with 370 atoms. It comprises MD trajectories for seven systems spanning four major classes of biomolecules and supramolecular structures.

\vspace{0.4cm}
\noindent \textit{Computational Methodology}:
All calculations were performed using the FHI-aims electronic structure software,\cite{blum2009ab} in conjunction with i-PI for the MD simulations. The potential energy and atomic force labels were determined at the PBE+MBD level of theory.\cite{seifert1996calculations,tkatchenko2012accurate} Two different types of basis sets were employed, referred to as "light" and "tight" within the FHI-aims framework. Trajectories were sampled at a resolution of 1 femtosecond (fs), with a thermostat used to regulate the temperature during simulations.

\vspace{0.4cm}
\noindent \textit{Data Accessibility}:
The MD22 data set is freely available for download at \url{http://www.sgdml.org/#data sets}.\cite{chmielamd22}

\subsection{WS22}
WS22\cite{pinheiro2023ws22} is a comprehensive database focusing on ten organic molecules, varying in size and complexity, with up to 22 atoms. It includes 1.18 million geometries, encompassing both equilibrium and non-equilibrium states. These geometries are meticulously sampled from specific distributions and further augmented with interpolated structures, resulting in a highly diverse data set. For each geometry, WS22 provides various QM properties, such as potential energies, forces, dipole moments, and electronic energies.

\vspace{0.4cm}
\noindent \textit{Computational Methodology}:
The construction of the WS22 database involves a meticulous four-step process. In the first step, researchers employed DFT to determine the equilibrium geometries for various conformations of each molecule. Calculations were conducted for both ground and excited electronic states (where applicable) to capture a broad range of configurations. All geometry optimizations were performed without symmetry constraints using the Gaussian~09 program with the hybrid density functional PBE0 and the 6-311G* basis set.\cite{krishnan1980self} For certain molecules, particularly those experiencing significant conformational changes upon excitation, first excited state (S$_1$) calculations were performed. In these cases, the linear-response time-dependent DFT approach was used for both geometry optimizations and frequency calculations at the same theoretical level, PBE0/6-311G*. Frequency calculations were then carried out to ensure that the optimized structures were true minima on the potential energy surface.

Beyond the initial equilibrium structures, the researchers implemented two techniques to create a diverse set of geometries. The first technique, Wigner sampling, considers the zero-point energy of molecules to generate a broader range of configurations compared to classical simulations.\cite{colomes2015comparing} This method uses the optimized geometries and harmonic frequencies obtained in step one to create an ensemble of non-equilibrium structures based on a Wigner probability distribution. For each molecule, 100,000 geometries were generated, ensuring a good distribution across different conformers.

To further explore beyond the vibrational degrees of freedom covered by Wigner sampling, researchers performed geodesic interpolation\cite{zhu2019geodesic} between every possible combination of stable conformers. This creates a smooth transition between different conformational spaces, including regions near transition states on the potential energy surface, which are not accessible via Wigner sampling. This technique added an additional 20,000 geometries per data set in the WS22 database.

In the final step, researchers conducted electronic structure calculations for all generated geometries, totaling 1.18 million calculations. This step assigns quantum chemical properties to each geometry in the database, including potential energies, atomic forces, dipole moments, and other relevant characteristics.

\vspace{0.4cm}
\noindent \textit{Data Accessibility}:
The WS22 database is publicly available on Zenodo at \url{https://doi.org/10.5281/zenodo.7032334}.\cite{pinheiro2023zenodo}  Each molecule in the database has its own file, containing information on all generated geometries: Wigner sampled geometries, interpolated geometries, and minimum energy conformations obtained via geometry optimization.

\subsection{VIB5}
The VIB5 database\cite{zhang2022vib5} is a meticulously curated collection of high-quality \textit{ab initio} quantum chemical data for five small polyatomic molecules with significant astrophysical relevance: methyl chloride (\ce{CH3Cl}), methane (\ce{CH4}), silane (\ce{SiH4}), fluoromethane (\ce{CH3F}), and sodium hydroxide (NaOH). The database includes over 300,000 grid points, with individual molecules featuring between 15,000 and 100,000 points. Each grid point corresponds to a specific nuclear configuration of the molecule, providing theoretical best estimates (TBEs) of potential energies and their constituent terms. Additionally, the data set offers energy and energy gradient calculations at various levels of theory, including MP2/cc-pVTZ and CCSD(T)/cc-pVQZ, as well as HF energies derived from these calculations using the corresponding basis sets.

\vspace{0.4cm}
\noindent \textit{Computational Methodology}:

The VIB5 database is built on a foundation of grid points that represent various nuclear configurations of the target molecules. These grid points were collected in VIB5 from the previous studies by some of the authors. Detailed descriptions of the grid point generation process can be found in the published VIB5 data descriptor\cite{zhang2022vib5} and authors' original publications.

For each molecule, TBEs and energy corrections were sourced directly from previous studies by the authors. Each TBE is calculated as the sum of several constituent terms, such as coupled-cluster energy at the complete basis set limit, core-valence electron correlation energy correction, and diagonal Born--Oppenheimer correction. Not all constituent terms were calculated at the same level of theory across all molecules; specific details are provided in Table 2 of the published VIB5 data descriptor. Various versions of MOLPRO and CFOUR programs\cite{werner2012molpro, werner2020molpro, matthews2020coupled} were utilized for these calculations.

In addition to TBEs, the VIB5 database includes complementary results from calculations performed at two distinct levels of theory: MP2/cc-pVTZ and CCSD(T)/cc-pVQZ, using the CFOUR program package (Versions 1.0 and 2.1). These calculations were conducted with multiple options, including FROZEN\_CORE = OFF (ON), SCF\_CONV = 10, CC\_CONV = 10, LINEQ\_CONV = 8, and CC\_PROG = ECC (VCC). From these calculations, HF energies were also extracted using the corresponding basis sets cc-pVTZ and cc-pVQZ. Additionally, for \ce{CH3Cl}, MP2/aug-cc-pVQZ energies were included, calculated using MOLPRO2012.

\vspace{0.4cm}
\noindent \textit{Data Accessibility}:

The VIB5 data is stored in a JSON file named VIB5.json, which can be downloaded from \url{https://doi.org/10.6084/m9.figshare.16903288}.\cite{vib5fig} The file structure organizes the data by molecule, providing information on chemical formula, number of atoms, and more. Each grid point entry includes nuclear positions (both Cartesian and internal coordinates) and various property values. These property values include energies (TBE, constituent terms, HF, MP2, CCSD(T)) and energy gradients (MP2, CCSD(T)). Specific JSON keys allow access to different properties for each grid point, as detailed in Table 3 of the original report.

\subsection{ANI-1}
The ANI-1 data set\cite{smith2017ani} provides a comprehensive collection of DFT total energy calculations of small organic molecules, encompassing approximately 20 million non-equilibrium conformations of 57,462 molecules. ANI-1 data set was used to train the general-purpose ANI-1 potential.\cite{smith2017anipotential}

\vspace{0.4cm}
\noindent \textit{Computational Methodology}:
The creation of the ANI-1 data set involved a multi-step process including molecule selection, geometry optimization, and normal mode sampling. Initially, the GDB-11 database\cite{fink2005virtual, fink2007virtual} was used as the source, containing organic molecules with up to 11 heavy atoms (C, N, O, F). A subset of 57,947 molecules with 1--8 heavy atoms limited to C, N, and O was chosen. RDKit software was employed to generate 3D conformations, ensuring neutral charge and singlet electronic ground state.

Next, each molecule underwent a pre-optimization using the MMFF94 force field.\cite{halgren1996merck}  This was followed by optimization at the $\omega$B97X/6-31G(d) level of theory\cite{ditchfield1971self} using the Gaussian 09 package to identify energy minima. Molecules that failed to converge were excluded, accounting for 0.84\% of the total data set.

Subsequently, normal mode coordinates and corresponding force constants were obtained for each optimized molecule. Random displacements for each mode were calculated based on temperature, and these displacements were used to generate new conformations by scaling the normalized normal mode coordinates. Single-point energy calculations were performed at the $\omega$B97X level for these new conformations. The number of conformations generated (N) depended on the number of heavy atoms (S) and the molecule's degrees of freedom (K).

\vspace{0.4cm}
\noindent \textit{Data Accessibility}:
The ANI-1 data set is stored in an HDF5 file accessible through Figshare at \url{https://doi.org/10.6084/m9.figshare.5287732.v1}. Additionally, a GitHub repository at \url{https://github.com/isayev/ANI1_dataset} provides usage details and data access instructions.\cite{anifiggit}

\subsection{ANI-1ccx and ANI-1x}
The ANI-1x and ANI-1ccx data sets\cite{smith2020ani} are foundational resources used to train the universal ANI-1x\cite{smith2018less} and ANI-1ccx\cite{smith2019approaching} ML potentials, respectively. These data sets provide an extensive collection of millions of organic molecule conformations containing the elements carbon, hydrogen, nitrogen, and oxygen (CHNO). The ANI-1x data set includes DFT calculations for approximately 5 million organic molecule conformations, obtained through an active learning algorithm. This data set features a much wider variety of molecular conformations compared to the ANI-1 random sampled data set, owing to the active learning methods employed in its construction. The ANI-1ccx data set is a carefully selected 10\% subset of the ANI-1x data set, recomputed with CCSD(T)/CBS level of theory.

\vspace{0.4cm}
\noindent \textit{Computational Methodology}:
The ANI-1x data set was developed using an active learning procedure to enhance the diversity and accuracy of the ANI-1x potential. Initially, an ensemble of ANI models was trained using bootstrapping on a preliminary data set. Molecules were then randomly sampled from extensive databases like GDB-11 and ChEMBL.\cite{fink2005virtual, fink2007virtual, mendez2019chembl} Four sampling techniques were applied to these molecules within active learning: MD sampling, normal mode sampling, dimer sampling, and torsion sampling. Data points with high uncertainty, identified using the ensemble disagreement measure $\rho$, were selected for DFT calculations. These high-uncertainty points were then added to the training set, and the models were retrained. This process was iteratively repeated to create a more diverse and comprehensive data set.

The ANI-1ccx data set leverages a higher level of theory, CCSD(T)*/CBS (special extrapolation scheme to CCSD(T)/complete basis set), for a subset of the ANI-1x data points. The selection process began with a random subset of ANI-1x data, for which CCSD(T)*/CBS reference data was generated. An ensemble of models was trained on this CCSD(T)*/CBS data, and $\rho$ was calculated for all remaining ANI-1x data points. Data points with $\rho$ values exceeding a predetermined threshold were then selected, and CCSD(T)*/CBS data for these points was generated and added to the training set. This iterative process continued until the final ANI-1ccx data set was established.

\vspace{0.4cm}
\noindent \textit{Data Accessibility}:
The ANI-1x and ANI-1ccx data sets are available in a single HDF5 file, accessible via Figshare at \url{https://doi.org/10.6084/m9.figshare.c.4712477}.\cite{smith2020figshare} A Python script (example\_loader.py) provided in a publicly accessible GitHub repository \url{https://github.com/aiqm/ANI1x\_data sets} can be used to access the data. The data set includes various properties such as energies (multiple types), forces, electronic multipole moments, and charges. These properties were calculated using three primary electronic structure methods: $\omega$B97X/6-31G*,\cite{ ditchfield1971self} $\omega$B97X/def2-TZVPP,\cite{weigend2005balanced} and CCSD(T)*/CBS. All $\omega$B97X/6-31G* calculations were performed using the Gaussian 09 electronic structure package, while $\omega$B97X/def2-TZVPP and CCSD(T)*/CBS calculations were conducted using the ORCA software package.\cite{neese2020orca}.


\red{

\subsection{ANI-2x}
The ANI-2x\cite{devereux2020extending} data set is used to train the ANI-2x ML model and contains 8.9 million nonequilibrium neutral singlet molecules with seven chemical elements (H, C, N, O, S, F, Cl) compared to its predecessor ANI-1x\cite{smith2020ani} with compounds comprising only four elements (H, C, N, O), thus enabling wider ability on drug discovery and biomolecules. Different from previous ANI datasets, ANI-2x uses S66x8 data set\cite{brauer2016s66x8}  to improve non-bonded interactions and a new type of sampling to improve bulk water.

\vspace{0.4cm}

\noindent \textit{Computational Methodology}:
ANI-2x data set is based on GDB-11\cite{fink2005virtual, fink2007virtual}, ChEMBL\cite{mendez2019chembl}, S66x8\cite{brauer2016s66x8} data sets, and additional randomly-generated amino acids and dipeptides to build the data sets. Similar active learning processes as previous ANI datasets are used to generate non-equilibrium geometries with refinement on torsion, non-bonded interaction, and bulk water. The obtained conformers are labeled with energies and forces at $\omega$B97X/6-31G* level of theory using Gaussian~09.  

\vspace{0.4cm}

\noindent \textit{Data Accessibility}:
The ANI-2x data sets are available on Zenodo with guidelines on use and methods used for generation \url{https://doi.org/10.5281/zenodo.10108942}.\cite{ani2zenodo} Except for the target level $\omega$B97X/6-31G* used to train the ANI-2x potential, additional reference data at 4 DFT levels are provided.

}

\subsection{Transition1x}
The Transition1x data set\cite{schreiner2022transition1x} is a unique resource for researchers developing ML models that can handle reactive systems. Unlike existing data sets that primarily focus on near-equilibrium configurations, Transition1x incorporates crucial data from transition regions. This empowers ML models to learn features essential for accurate reaction barrier prediction. Furthermore, Transition1x serves as a new benchmark for evaluating how well ML models capture reaction dynamics.

With 9.6 million data points, each meticulously generated using DFT calculations, Transition1x encompasses the forces and energies for a staggering 10,000 organic reactions. The core data generation method is the Nudged Elastic Band (NEB)\cite{sheppard2008optimization} technique which efficiently explores millions of configurations within transition state regions, providing a comprehensive picture of the energetic landscape during reactions.

\vspace{0.4cm}

\noindent \textit{Computational Methodology}:
The creation of Transition1x follows a meticulous workflow. First, a comprehensive database of reactant-product pairs for various organic reactions lays the foundation, providing initial geometries for reactants, products, and transition states.\cite{grambow2020reactants} To ensure compatibility with the popular ANI1x data set,\cite{smith2018less} Transition1x utilizes the same $\omega$B97X functional\cite{chai2008systematic} and the 6-31G(d) basis set\cite{ditchfield1971self} for DFT calculations. These calculations are performed using the ORCA software.\cite{neese2020orca}

Next comes the NEB\cite{sheppard2008optimization} and CINEB\cite{henkelman2000climbing} optimization stage. NEB, along with CINEB, plays a central role. Both techniques are employed alongside the BFGS optimizer to efficiently refine the initial guess for the minimum energy pathway (MEP). The workflow involves relaxing reactant and product geometries, constructing an initial path, minimizing its energy using image dependent pair potential (IDPP)\cite{smidstrup2014improved} within the NEB framework, and finally using NEB and CINEB together to achieve an accurate MEP.

To ensure high-quality data, filtering steps are crucial. Unphysical configurations and redundant data points are discarded. Reactions that fail to converge during NEB calculations are excluded, and only reaction paths that significantly differ from previously included ones are incorporated.

\vspace{0.4cm}

\noindent \textit{Data Accessibility}:
The Transition1x data set is readily available on Figshare at \url{https://doi.org/10.6084/m9.figshare.19614657.v4} and stored in a single, user-friendly HDF5 file named "Transition1x.h5".\cite{transition1xfigshare} Dataloaders and example scripts are available at \url{https://gitlab.com/matschreiner/T1x}.

\subsection{QM-sym}
Symmetry plays a pivotal role in determining various molecular characteristics, such as excitation degeneracy and transition selection rules. Despite its significance, many quantum chemistry databases often overlook this crucial aspect. Bridging this gap, the QM-sym database\cite{liang2019qm} emerges as a valuable resource, meticulously documenting the C$_{n\text{h}}$ symmetry for each molecule within its vast repository. Comprising 135,000 structures featuring elements like H, B, C, N, O, F, Cl, and Br, it encompasses symmetries like C$_{2\text{h}}$, C$_{3\text{h}}$, and C$_{4\text{h}}$.

\vspace{0.4cm}

\noindent \textit{Computational Methodology}:
QM-sym employs a two-step process, leveraging symmetry to streamline database generation. Initially, it constructs fundamental molecular frameworks based on standardized bond lengths and angles. Subsequently, employing a genetic algorithm, these structures undergo iterative refinement to attain stable configurations while conforming to predefined symmetrical point groups (e.g., C$_{2\text{h}}$ or D$_{6\text{h}}$).

Following structural generation, each molecule is optimized at the B3LYP/6-31G(2df,p) level theory with Gaussian 09 software, which ensures chemical validity and stability. However, complexities may lead to convergence failures or unstable configurations. To mitigate this, a filtering mechanism akin to QM9 is enacted, setting a maximum optimization cycle limit of 200 and implementing stricter convergence criteria. Ultimately, only structures exhibiting successful optimization and positive vibrational frequencies, indicative of stability, are retained in the final QM-sym database, striking a balance between symmetry incorporation and molecular stability.

\vspace{0.4cm}

\noindent \textit{Data Accessibility}:
The QM-sym database is readily accessible through platforms like GitHub and Figshare using the links \url{https://github.com/XI-Lab/QM-sym-database} and \url{https://doi.org/10.6084/m9.Figshare.9638093}, respectively.\cite{Liang2019git, liang2019fig} It provides QM\_sym.xyz files containing atomic coordinates and a plethora of predicted properties, encompassing point group information, enthalpies, atomization energies, zero-point energies, and energy and symmetry labels spanning from HOMO$-$5 to LUMO+5. Each structure is uniquely indexed by QM\_sym\_i.xyz, with 'i' denoting the structure's database order.

\subsection{QM-symex}
While QM-sym focuses on larger, symmetrical organic molecules and incorporates vital symmetry information, it lacks crucial excited-state properties necessary for applications like photosensitizers and PDT.

To fill this gap, the team behind QM-sym introduced QM-symex,\cite{liang2020qm} a database dedicated to excited-state information. In contrast to existing databases such as QM8,\cite{ruddigkeit2012enumeration,ramakrishnan2015electronic} QM-symex offers detailed insights into critical parameters like oscillator strength, transition energy, and transition symmetry.

\vspace{0.4cm}

\noindent \textit{Computational Methodology}:
QM-symex inherits the rigorous data generation process established in QM-sym, ensuring the stability and symmetry of its molecules. Building upon the 135,000 structures from QM-sym, an additional 38,000 molecules were meticulously generated. Maintaining symmetry preservation and stability remains a core principle. Using Gaussian 09 optimization with 100 cycles, each molecule undergoes rigorous validation to retain its original symmetry throughout the process. 

\vspace{0.4cm}

\noindent \textit{Data Accessibility}:
Accessible on Figshare at \url{https://doi.org/10.6084/m9.Figshare.12815276},\cite{Liang2020fig} QM-symex provides data in XYZ format, indexed as QM\_symex\_i.xyz for ease of access,where 'i' represents the order of the structure in the database. In addition to inheriting all properties from QM-sym, QM-symex enriches the data set with information on the first ten singlet and triplet transitions. This includes details such as energy, wavelength, orbital symmetry, transition distance, and other quasi-molecular properties. With the new 38k molecules, the distribution of symmetries within QM-symex is noteworthy. C$_{2\text{h}}$ symmetry occupies a significant proportion of 46\%, while C$_{3\text{h}}$ and C$_{4\text{h}}$ symmetries account for 41\% and 13\%, respectively. 

\subsection{$\nabla^2$DFT}
The $\nabla^2$DFT\cite{khrabrov2024nabla} data set provides a comprehensive collection of approximately 16 million conformers for around 2 million drug-like molecules, featuring 8 atom types (H, C, N, O, Cl, F, Br) and up to 62 atoms. It is an extension of the original $\nabla$DFT data set\cite{khrabrov2022nabladft}, offering energies, forces, and various other properties calculated at a reasonably accurate DFT level for a wide range of molecules. Additionally, the data set includes relaxation trajectories for numerous drug-like molecules. Based on the Molecular Sets (MOSES) data set\cite{polykovskiy2020molecular}, $\nabla^2$DFT contains around 1.93 million molecules with atoms C, N, S, O, F, Cl, Br, and H, 448,854 unique Bemis-Murcko scaffolds\cite{bemis1996properties}, and 58,315 unique BRICS fragments\cite{degen2008art}.

\vspace{0.4cm}

\noindent \textit{Computational Methodology}:
The $\nabla^2$DFT data set employs a meticulous approach to generate a diverse set of conformations (1 to 100) for each molecule. Initially, the conformation generation method from the RDKit software suite was utilized.

To avoid redundancy, a clustering step was implemented using the Butina clustering method,\cite{barnard1992clustering} which groups similar conformations based on their geometrical properties. This step effectively reduced the number of conformations while ensuring a representative set is retained. Only clusters that encompassed at least 95\% of the conformations for a specific molecule were kept. The most central conformation within each retained cluster, known as the centroid, was chosen as the representative conformation for that molecule in the final data set. The resulting data set contained 1 to 69 unique conformations per molecule, totaling over 12 million conformations.

For each conformation, electronic properties (energy, DFT Hamiltonian matrix, DFT overlap matrix, etc.) were computed using the Kohn--Sham DFT method at the $\omega$B97X-D/def2-SVP level of theory with the Psi4 quantum chemistry software. Additionally, interatomic forces were calculated at the same level of theory for about 452,000 molecules and 2.9 million conformations.

\vspace{0.4cm}

\noindent \textit{Data Accessibility}:
The associated code and links to access various splits of the $\nabla^2$DFT data set, are freely available for download at \url{https://github.com/AIRI-Institute/nablaDFT}.

\red{

\subsection{The COMPAS Project}
The COMPAS (COMputational database of Polycylic Aromatic Systems) project aims at providing structures and properties for ground-state polycyclic aromatic systems. COMPAS-1\cite{wahab2022compas} focuses on cata-condensed polybenzenoid hydrocarbons (PBHS) with 43k molecules, COMPAS-2\cite{mayo2024compas} provides 0.5 million cata-condensed poly(hetero)cyclic aromatic molecules and COMPAS-3\cite{wahab2024compas} explores 40k the peri-condensed polybenzenoid hydrocarbons. All the 3 data sets contain molecules up to 11 rings with properties calculated at xTB level and molecules up to 10 rings at DFT level.

\vspace{0.4cm}

\noindent \textit{Computational Methodology}:
For COMPAS-1, the initial polycyclic molecules are constructed with CaGe\cite{brinkmann2010gage} up to 11 rings. All the geometries were optimized at GFN2-xTB level and molecules containing up to 10 rings were further optimized at B3LYP-D3(BJ)/def2-SVP with ORCA 4.2.0. COMPAS-2 starts from various building blocks different in size and elements contained with the final geometries optimized using GFN1-xTB and CAM-B3LYP-D3BJ/def2-SVP. COMPAS-3 follows the same protocol as COMPAS-1 and perform calculation on the obtained molecules at GFN2-xTB and CAM-B3LYP-D3BJ/aug-cc-pVDZ//CAM-B3LYP-D3BJ/def2-SVP level of theory.

\vspace{0.4cm}

\noindent \textit{Data Accessibility}:
COMPAS datasets are publicly available at \url{https://gitlab.com/porannegroup/compas} where molecules and their corresponding QM properties including electronic energies, HOMO/LUMO energies and gap, ZPE, etc., are hosted. An online website is developed for easier structure search and download at \url{https://compas.net.technion.ac.il/}.\cite{compas2022availability}
}

\red{
\subsection{CREMP}
CREMP (Conformer-Rotamer Ensembles of Macrocyclic Peptides) data set\cite{grambow2024cremp} aims to explore macrocyclic peptides which is lacking in MLP field, providing 36k representative 4-, 5-, and 6-mer homodetic cyclic peptides with up to 31.3 million unique conformers with properties calculated at semiempirical level.  CREMP-CycPeptMPDB data set is also provided which is built based on the established CycPeptMPDB\cite{jiannan2023cycpeptmpdb} database featuring passive membrane permeability with 8.7 million newly-generated conformers from 3k 6-, 7-, and 10-mer cyclic peptides.
\vspace{0.4cm}

\noindent \textit{Computational Methodology}:
To generate the CREMP data set, we utilized the CREST tool to explore diverse conformational ensembles of macrocyclic peptides. We began with a set of 36,198 unique homodetic macrocyclic peptides, sampled based on key parameters such as side chains, stereochemistry, and N-methylation. Using RDKit,\cite{rdkitsoft} each peptide was converted to canonical SMILES, and initial conformers were generated via the ETKDGv3 method,\cite{riniker2015better} followed by energy optimization using the MMFF94 force field.\cite{halgren1996merck} The 1,000 lowest-energy conformers were further optimized using GFN2-xTB\cite{bannwarth2021extended} in chloroform solvent. These optimized structures were then used as inputs for CREST simulations,\cite{pracht2020automated} where 14 metadynamics runs were performed in parallel. The final ensembles were filtered to ensure chemical graph consistency across all conformers.
\vspace{0.4cm}

\noindent \textit{Data Accessibility}:
CREMP and CERMP-CycPeptMPDB data sets are available on Zenodo at \url{https://doi.org/10.5281/zenodo.7931444} and \url{https://doi.org/10.5281/zenodo.10798261}.\cite{crempzenodo} The provided data consists of CREST metadata including energies, entropies, etc., and conformers with their 3D strucutures. CREPM also provides instructions of data usage on GitHub at \url{https://github.com/Genentech/cremp}.

\subsection{GEOM}
The GEOM ( Geometric Ensemble Of Molecules) data set\cite{axelrod2022geom} is a comprehensive resource containing high-quality conformers for 451,186 organic molecules. This includes 317,928 drug-like species with experimental data and 133,258 molecules from the QM9 data set.\cite{ramakrishnan2014quantum} The dataset encompasses 304,466 molecules sourced from AICures (\url{https:// www.aicures.mit.edu}), a machine learning challenge focused on drug repurposing for COVID-19 and related illnesses, as well as 16,865 molecules from the MoleculeNet benchmark,\cite{zhenqin2018molnet} annotated with experimental properties relevant to physical chemistry, biophysics, and physiology. 

\vspace{0.4cm}

\noindent \textit{Computational Methodology}:
The data generation process for the GEOM data set involved several key steps. It began with SMILES pre-processing, where SMILES strings were converted to their canonical forms using RDKit to ensure consistency across the dataset. For drug molecules found in clusters (e.g., with counter-ions or salts), de-salting procedures were applied to isolate the main compound and adjust its ionization state. This step ensured uniform molecular representation across multiple sources.

In the initial structure generation phase, RDKit was used to generate molecular conformers, which were optimized using the MMFF force field. The ten lowest energy conformers were further optimized with GFN2-xTB, and the lowest energy conformer was selected as the starting structure for the CREST simulations.\cite{pracht2020automated} For the QM9 dataset, molecules were re-optimized with xTB before undergoing CREST, despite already being optimized with DFT.

The next step involved graph re-identification, which was crucial for handling changes in stereochemistry and reactivity during CREST simulations. RDKit was used to assign graph features to each conformer, ensuring accuracy in molecular structures. For 534 molecules from the BACE dataset, CENSO simulations\cite{grimme2021censo} were performed to further refine conformers with DFT geometry optimizations.

Finally, with ORCA version 5.0.2,  single-point DFT calculations (with r2scan-3c functional, mTZVPP basis, C-PCM model of water, and default grid 2) were carried out on all CREST conformers from the BACE dataset to provide high-accuracy energies. Additionally, Hessian calculations were performed using xTB to obtain vibrational frequencies, providing valuable thermodynamic information.

\vspace{0.4cm}

\noindent \textit{Data Accessibility}:
The GEOM is online at \url{https://github.com/learningmatter-mit/geom}, where both source datasets and tutorials on loading and analyzing data are provided. Conformer-level information (geometry, energy, degeneracy, etc.) and species-level information (experimental values, average conformer energy, etc.) are included.\cite{geom2022github}

\subsection{tmQM}
The tmQM data set\cite{balcells2020tmqm} explores transition metal-organic compound space including Werner, bioinorganic and organometallic complexes with their respective QM properties based on Cambridge structural database,\cite{groom2016cambridge} enabling wide coverage of organometallic space. There are in total 86,665 complexes with 3d, 4d, and 5d transition metals from groups 3 to 12. Lately, 60k graph data set (tmQMg)\cite{kneiding2023tmqmg} and a 30k ligand library (tmQMg-L)\cite{kneiding2024tmqmgl} are constructed with NBO analysis on tmQM dataset.

tmQM offers Cartesian coordinates optimized at the GFN2-xTB level, along with quantum properties computed at the DFT (TPSSh-D3BJ/def2-SVP) level. These properties include electronic and dispersion energies, metal center natural charge, HOMO/LUMO energies and gaps, dipole moments, and polarizabilities calculated at the GFN2-xTB level.

\vspace{0.4cm}

\noindent \textit{Computational Methodology}:
The tmQM data set was generated by extracting structures from the 2020 release of the Cambridge Structural Database (CSD) using seven filters: 1) inclusion of only mononuclear transition metal (TM) compounds, 2) containing at least one carbon and one hydrogen, 3) excluding non-metal components, 4) omitting polymeric structures, 5) ensuring 3D coordinates, 6) excluding disordered atoms, and 7) removing highly charged species (charge outside -1 to +1). From the filtered CSD data, 116,332 structures were selected. After geometry optimization using the GFN2-xTB method, additional filters for convergence, geometry quality, and electron count reduced the dataset to 86,699 structures. Ultimately, quantum mechanical properties were computed using DFT with the hybrid meta-GGA TPSSh functional and the def2-SVP basis set through Gaussian 16.

\vspace{0.4cm}

\noindent \textit{Data Accessibility}:
The tmQM dataset and its derivatives are available at \url{https://github.com/bbskjelstad/tmqm} with usage information as well as timely updates. It is also part of the Quantum-Machine project and can be found at \url{http://quantum-machine.org/datasets/}.

\subsection{OFF-ON dataset}
The OFF-ON (organic fragments from organocatalysts that are non-modular) database\cite{celerse2024offon} is developed to train machine learning models aimed at predicting the behavior of structurally and conformationally diverse functional organic molecules, particularly photoswitchable organocatalysts. This dataset contains 7,869 equilibrium geometries and 67,457 non-equilibrium geometries of organic compounds and dimers, providing a broad foundation for exploring flexible molecular structures and their free energy surfaces. 

\vspace{0.4cm}

\noindent \textit{Computational Methodology}:
The data generation process for the OFF-ON database involved two key stages. The first stage concentrated on creating a set of molecules that represent diverse chemical environments pertinent to functional organic molecules. This set included catalytic moieties, photochromic units, substituted aromatic rings, and representations of non-covalent interactions. To compile this diverse set, existing molecular databases such as OSCAR, CSD, and PubChem\cite{kim2023pubchem} were utilized, alongside additional generation protocols detailed in the supplementary materials. As a result, a final collection of 7,869 unique entries was established, encompassing 3,533 catalytic moieties, 538 photochromic units, 3,165 substituted rings, and 633 dimers representing non-covalent interactions.

The second stage focused on generating a range of molecular conformations to capture out-of-equilibrium effects. This was achieved through molecular dynamics (MD) simulations, where for each of the 7,869 structures, 5 ps MD trajectories were initiated from geometries optimized using DFTB. These simulations resulted in over 2 million new geometries. The dataset was subsequently refined using the farthest point sampling (FPS) algorithm, which selected the most diverse conformations from the MD trajectories. This process led to the inclusion of 67,457 unique out-of-equilibrium structures in the database.

In total, the OFF-ON database comprises 75,326 structures, combining both equilibrium and non-equilibrium states. The computational details of the dataset include electronic structure computations and machine learning potential modeling. The baseline energy calculations were performed at the DFTB3 level with the 3ob parameters and D3 dispersion correction, implemented in the DFTB+ software. These results were normalized using multilinear regression models. The reference energy was computed at the PBE0-D3/def2-SVP level using the TeraChem software, and the difference between DFTB and PBE0 energies was used as a correction.

\vspace{0.4cm}

\noindent \textit{Data Accessibility}:
The OFF-ON database is openly available at \url{https://archive.materialscloud.org/record/2023.189}.\cite{offon2024availablity}

}

\subsection{Others}

In addition to the data sets and databases described above, this subsection introduces six more data sets and databases: C7O2H10-17,\cite{schutt2017quantum} ISO17,\cite{schutt2017schnet}, VQM24,\cite{khan2024towards}, QCDGE,\cite{zhu2024quantum} QM-22,\cite{bowman2022md17} CheMFi\cite{vinod2024chemfi}, QM9S\cite{zou2023qm9s} and the TensorMol ChemSpider data set.\cite{yao2018tensormol} Among them, the C7O2H10-17 data set comprises MD trajectories for 113 randomly selected isomers of C7O2H10. These trajectories were calculated at a temperature of 500 K with a high temporal resolution of 0.5~fs, using DFT with the PBE exchange-correlation potential. Notably, C7O2H10 represents the largest set of isomers within the QM9 data set. The identifiers used in this data set are consistent with those used in the QM9 isomer subset, ensuring compatibility and ease of reference.

The ISO17 data set extends the C7O2H10-17 data set, consisting of 129 isomers with the chemical formula C7O2H10. Each isomer includes 5,000 conformational geometries, energies, and forces, sampled at a resolution of 1 femtosecond in the MD trajectories. These simulations were conducted using the FHI-aims software, employing DFT with the PBE functional and the Tkatchenko-Scheffler (TS) van der Waals correction method.

The VQM24 data set is a comprehensive data set containing QM properties of small organic and inorganic molecules. It encompasses 258,242 unique constitutional isomers and 577,705 conformers of varying stoichiometries, focusing on molecules composed of up to five heavy atoms from C, N, O, F, Si, P, S, Cl, Br. For each molecule, the data set provides optimized structures, thermal properties (vibrational modes, frequencies), electronic properties, wavefunctions, and, for a subset of 10,793 conformers, diffusion quantum monte carlo (DMC) energies. Calculations were performed at the $\omega$B97X-D3/cc-pVDZ level of theory, with conformers initially generated using GFN2-xTB and subsequently relaxed using DFT. Calculations were performed with the following computational programs: Surge,\cite{mckay2022surge} RDKit,\cite{rdkitsoft} Crest,\cite{pracht2020automated} Psi4\cite{smith2020psi4} and QMCPACK.\cite{kent2020qmcpack} 

The QCDGE database offers an extensive collection of ground and excited-state properties for 443,106 organic molecules, each containing up to ten heavy atoms, including carbon, nitrogen, oxygen, and fluorine. These molecules are sourced from well-known databases such as QM9, PubChemQC, and GDB-11. The database provides 27 molecular properties, including ground-state energies, thermal properties, and transition electric dipole moments, among others. Ground-state geometry optimizations and frequency calculations for all compounds were carried out using the B3LYP/6-31G* level of theory with BJD3 dispersion correction,\cite{grimme2011effect} while excited-state single-point calculations were performed at the $\omega$B97X-D/6-31G* level. All computational work was conducted using Gaussian 16.

The QM-22 database is a compilation of molecular data sets specifically curated for DMC calculations of the zero-point state. Each data set within QM22 employs unique methodologies tailored to the specific molecules involved. Detailed computational methods for each data set can be found in their corresponding publications.\cite{bowman2022md17}

The CheMFi data set is a multifidelity compilation of quantum chemical properties derived from a subset of the WS22 database, featuring 135,000 geometries sampled from nine distinct molecules. CheMFi encompasses five different levels of fidelity, each corresponding to a specific basis set size (STO-3G, 3-21G, 6-31G, def2-SVP, def2-TZVP). The data set was generated using TD-DFT calculations with the CAM-B3LYP functional, performed via the ORCA software package. It includes comprehensive data on properties such as vertical excitation energies, oscillator strengths, molecular dipole moments, and ground state energies. Moreover, computational times for each fidelity level are provided to facilitate benchmarking of multifidelity models.

\red{The QM9S dataset, used for training and testing the DetaNet deep learning model, consists of 133,885 organic molecules derived from the QM9 dataset. The molecular geometries were re-optimized using the Gaussian~16 software at the B3LYP/def-TZVP level of theory. A wide range of molecular properties were then calculated at the same level, including scalar values (e.g., energy, partial charges), vectors (e.g., electric dipole), second-order tensors (e.g., Hessian matrix, quadrupole moment, polarizability), and third-order tensors (e.g., octupole moment, first hyperpolarizability). Additionally, frequency analysis and time-dependent DFT were performed to obtain the IR, Raman, and UV-Vis spectra.
}

The TensorMol ChemSpider data set contains potential energies and forces for 3 million conformations from 15,000 different molecules composed of carbon, hydrogen, nitrogen, and oxygen. These molecules were sourced from the ChemSpider database.\cite{pence2010chemspider} The training geometries were sampled using metadynamics,\cite{herr2018metadynamics} and their energies were calculated using the QChem program\cite{shao2015advances} with the $\omega$B97X-D exchange-correlation functional and a 6-311G** basis set.

\noindent \textit{Data Accessibility}:
The C7O2H10-17 and ISO17 data sets are available at \url{http://quantum-machine.org/datasets/}, while the VQM24 data set can be accessed on Zenodo at \url{https://doi.org/10.5281/zenodo.11164951}. The QCDGE database is hosted at \url{http://langroup.site/QCDGE/}, and the QM22 database can be downloaded from \url{https://github.com/jmbowma/QM-22}. The CheMFi data set is available at \url{https://github.com/SM4DA/CheMFi}. The QM9S data set is hosted at Figshare\url{https://doi.org/10.6084/m9.figshare.24235333}. Although the TensorMol ChemSpider data set was once accessible via \url{https://drive.google.com/drive/folders/1IfWPs7i5kfmErIRyuhGv95dSVtNFo0e_}, as mentioned in the supplementary information, it is no longer available.






\section{Concluding Remarks and Future Outlook}

The exploration of quantum chemical databases and data sets for ML potentials has revealed a rich landscape of resources offering valuable data for training and validating these powerful tools. We have identified a diverse range of data sets and databases, each with its own strengths in terms of information content, level of theory, molecular coverage, and creation procedures.

As the number of quantum chemical data sets continues to grow rapidly, the importance of maintaining an up-to-date overview becomes critical. To this end, we provide an updatable resource (accessible at \url{https://github.com/Arif-PhyChem/datasets_and_databases_4_MLPs}), which aims to track and categorize emerging databases and data sets as they become available, ensuring researchers have access to the latest information. \red{This resource also comes with the machine-readable representation of the overview of the data sets, which allows to easier sort and filter the data sets when required. The corresponding Jupyter notebook is also provided.}

However, the field faces significant challenges in ensuring the long-term accessibility of these data sets. Instances like the disappearance of the TensorMol database highlight the fragility of data availability over time. Platforms such as Figshare and Zenodo offer solutions to some extent, with Zenodo providing free hosting for large data sets, while Figshare's service comes with a fee for larger storage needs. Ensuring the longevity and accessibility of data sets will require continued support and development of such repositories.

Moreover, data format standardization is another area requiring attention. The FAIR (Findable, Accessible, Interoperable, Reusable) principles provide a guiding framework,\cite{FAIR} but the diversity of formats in use today makes unification challenging. Efforts to promote standard formats, such as ioChem-BD\cite{ioChem-BD} and Quantum Chemistry Schema,\cite{QCSchema} could help streamline data exchange and enhance interoperability across different platforms and research groups. Attempts to provide standard formats interoperable between software are naturally undertaken in the big software ecosystems gluing together different quantum chemical and/or ML codes like is done in MLatom\cite{mlatom3} and ASE.\cite{larsen2017atomic}

\red{Another big issue with the growing number of data sets and databases is that they are often generated at different levels severely hampering their use in ML potentials as, e.g., training a ML potential on a merged data set would be very problematic if possible at all to achieve any reasonable result. One of the solutions is recalculating properties at the same level for the merged data sets but it is very computationally costly undertaking. Recently, some of us suggested a potentially simple solution to this problem by creating all-in-one machine learning potentials which take in input the level of theory as an additional feature: we showed that this approach allows to train a single, accurate ML model on a heterogeneous data containing data at different levels of theory.\cite{allinone24} This seems to be one of the most promising direction for the future explorations which we also pursue in our labs.}

Furthermore, maintaining updatable data sets and databases, particularly for "living" (i.e., updatable) methods like UAIQM,\cite{UAIQM} is crucial for the continued evolution of ML potentials. Ensuring these data sets and databases are regularly updated and accessible will require coordinated efforts within the research community, alongside the development of tools and standards that facilitate easy data curation and integration.

Finally, materials databases such as ChemSpider,\cite{pence2010chemspider} the Cambridge structural database,\cite{groom2016cambridge} and the materials project\cite{jain2020materials} serve as exemplary models of online accessibility, demonstrating how well-curated and easily navigable platforms can support a wide range of research activities. By drawing inspiration from these platforms, the quantum chemistry community can develop databases that are not only rich in content but also user-friendly and sustainable.

In conclusion, by addressing these challenges--ranging from the rapid proliferation of data sets to the need for standardized formats and long-term accessibility--the future of quantum chemical databases and data sets holds immense promise. The integration of active learning, high-throughput calculations, and other advanced methodologies will further enhance the utility of these resources, enabling ML potentials to reach their full potential in computational chemistry. This synergy between data and algorithms will accelerate scientific discovery, driving progress across a wide array of fields and deepening our understanding of the molecular world.

\section*{Acknowledgements}

A.U. acknowledges funding from the National Natural Science Foundation of China (No. W2433037). P.O.D. acknowledges support from the National Natural Science Foundation of China (funding via the Outstanding Youth Scholars (Overseas, 2021) project) and via the Lab project of the State Key Laboratory of Physical Chemistry of Solid Surfaces.

\bibliography{main}

\end{document}